# Graphene-based sponges for electrochemical degradation of persistent organic contaminants

Luis Baptista-Pires[†,‡], Giannis-Florjan Norra[†,‡], Jelena Radjenovic [†,§,*]

*Catalan Institute for Water Research (ICRA), Emili Grahit 101, 17003 Girona, Spain*

[‡]*University of Girona, Girona, Spain*

[§]*Catalan Institution for Research and Advanced Studies (ICREA), Passeig Lluís Companys 23, 08010 Barcelona, Spain*

* *Corresponding author:*

*Jelena Radjenovic, Catalan Institute for Water Research (ICRA), Emili Grahit 101, 17003 Girona, Spain*

Phone: + 34 972 18 33 80; Fax: +34 972 18 32 48; E-mail: jradjenovic@icra.cat

2

# Abstract

Graphene-based sponges doped with atomic nitrogen and boron were applied for the electrochemical degradation of persistent organic contaminants in one-pass, flow-through mode, and in a low-conductivity supporting electrolyte. The B-doped anode and N-doped cathode was capable of >90% contaminant removal at the geometric anodic current density of 173 A m$^{-2}$. The electrochemical degradation of contaminants was achieved via the direct electron transfer, the anodically formed $O_3$, and by the OH$^{\bullet}$ radicals formed by the decomposition of $H_2O_2$ produced at the cathode. The identified transformation products of iopromide show that the anodic cleavage of all three C-I bonds at the aromatic ring was preferential over scissions at the alkyl side chains, suggesting a determining role of the π- π interactions with the graphene surface. In the presence of 20 mM sodium chloride (NaCl), the current efficiency for chlorine production was <0.04%, and there was no chlorate and perchlorate formation, demonstrating a very low electrocatalytic activity of the graphene-based sponge anode towards chloride. Graphene-based sponges were produced using a low-cost, bottom-up method that allows easy introduction of dopants and functionalization of the reduced graphene oxide coating, and thus tailoring of the material for the removal of specific contaminants.

# Introduction

The ongoing water crisis caused by the water scarcity and pollution is inducing a change in the water management; the implementation of decentralized water and wastewater treatment, and the use of alternative water resources are being increasingly considered as part of the solution. Electrochemical processes have the potential to develop into platform technologies for decentralized (waste)water treatment as they do not use chemical reagents, do not form a residual waste stream, operate at ambient temperature and pressure, and are robust, versatile processes with small footprints and modular designs. However, a major drawback of electrochemical water treatment is rapid oxidation of chloride to chlorine (HOCl/OCl$^-$), which reacts with the organic matter to form toxic and persistent chlorinated byproducts (Radjenovic and Sedlak 2015). Chloride can be further oxidized to chlorate (ClO$_3^-$) and perchlorate (ClO$_4^-$) at highly oxidizing boron-doped diamond (BDD) (Bergmann et al. 2009, Garcia-Segura et al. 2015), and Magnéli-phase titanium suboxide (Ti$_4$O$_7$) anodes (Lin et al. 2018, Lin et al. 2020). Given that chloride is a major anion found in all natural waters, the formation of chlorinated byproducts represents a major challenge for the safe application of electrochemical water treatment systems.

Recent research efforts have been directed towards the development of porous electrodes based on carbon nanotubes (CNTs) and Ti$_4$O$_7$, which were demonstrated to electrochemically degrade a range of persistent organic contaminants (e.g., antibiotics, pesticides, dyes, phenols) (Gayen et al. 2018, Le et al. 2019, Liu et al. 2015, Vecitis et al. 2011). Nevertheless, the quest for a low-cost, electroactive material that can be synthesized in porous geometries is far from over. Apart from being electrocatalytically efficient, mechanically robust, and electrochemically stable, ideal electrode material should be produced by an easily scalable, low-cost method. Furthermore, it should have a low tendency to produce chlorinated byproducts. Three-dimensional (3D) graphene-based macrostructures were intensively studied as electrodes for supercapacitors, batteries, and sensors due to their high specific surface area and versatile surface chemistry (Raccichini et al. 2015, Yousefi et al. 2019). An attractive feature of graphene-based materials is the possibility to fine-tune their electrical (e.g., conductivity, electrocatalytic activity), physico-chemical (e.g., hydrophobicity, surface roughness) and mechanical properties (e.g., flexibility) by introducing defects and dopants into the graphene network. In water treatment, 3D graphene-based sponges and foams were applied for the removal of heavy metals, dyes and other organic pollutants from water via adsorption through electrostatic interactions and π–π interactions (Ali et al. 2019). Recent studies focused on the use of reduced graphene oxide (RGO)-coated graphite and carbon felts as cathodes for the *in situ* production of $H_2O_2$ via the $2e^-$ reduction of oxygen (Kim et al. 2018, Le et al. 2015, Mousset et al. 2016). The N-doping of graphene was reported to significantly enhance the cathodic production of $H_2O_2$ (Qu et al. 2010, Su et al. 2019, Yang et al. 2018, Yang et al. 2019). Moreover, pyrrolic-N is considered to enhance the $2e^-$ oxygen reduction to $H_2O_2$ (Li et al. 2020), and pyridinic-N groups were correlated with the *in situ* activation of $H_2O_2$ to hydroxyl radical (OH•) (Su et al. 2019). In addition, recent study reported the synergy between the pyrrolic-N- and B-dopants in the B,N co-doped carbon materials for the anodic generation of ozone ($O_3$) at the mesoporous carbon electrode (Zhang et al. 2020). Yet, 3D graphene-based macrostructures produced with graphite and carbon felts as supports are not well-suited for the anodic oxidation of pollutants due to the rapid corrosion of graphite and carbon already at potentials ≥0.207 V/SHE (vs Standard Hydrogen Electrode) (Eifert et al. 2020). Furthermore, template-free 3D graphene-based networks have very limited applications as electrode materials due to their extreme brittleness and difficult scale-up. Recently, mineral wool was proposed as structurally stable, low-cost template for the synthesis of 3D graphene macrostructures, and demonstrated excellent thermal stability during Joule heating to temperatures as high as 350ºC, applied for oil clean-up from water (Ge et al. 2017).

In this work, we synthesized B- and N-doped graphene-based sponges using a novel, low-cost hydrothermal self-assembly method based on the mineral wool as supporting template. The B-doped graphene-based sponge anode and N-doped graphene-based sponge cathode was employed for the electrochemical water treatment in a one-pass, flow-through reactor. We selected a set of organic contaminants known to be persistent to chemical and electrochemical oxidation/reduction, such as iodinated contrast media iopromide (IPM) and diatrizoate (DTR), antibacterial agent triclosan (TCL), and non-steroidal anti-inflammatory drug diclofenac (DCF) (Dickenson et al. 2009, Sein et al. 2008). To evaluate the system performance under the ohmic drop of real contaminated water, all experiments were conducted using a low-conductivity supporting electrolyte (~1 mS cm$^{-1}$) because most of the groundwater, surface water, and even municipal wastewater effluents are low conductivity solutions (<2 mS cm$^{-1}$). We investigated the impact of the applied current densities, flow rates and different flow directions on the removal of contaminants. In addition, the impact of chloride was studied in the presence of 20 mM NaCl. Finally, to gain insight into the degradation mechanisms occurring at the newly developed graphene-based sponge electrodes, we identified the major transformation products of IPM.

## Materials and methods

**Graphene-based sponge fabrication**

GO was provided from Graphenea S.L. Boric acid and urea were purchased from Sigma Aldrich. All chemicals and standards of target contaminants were of analytical grade. Mineral wool was purchased from Diaterm, Spain. Three different GO solutions were prepared: *i)* 4 g L$^{-1}$ GO solution to produce reduced graphene oxide (RGO) sponge, *ii)* 6 g of boric acid dissolved in 4 g L$^{-1}$ GO solution to produce B-doped reduced graphene oxide (BRGO) sponge, and *iii)* 42 g of urea dissolved in 4 g L$^{-1}$ GO solution to produce N-doped reduced graphene oxide (NRGO) sponge. The procedure of graphene-based sponge synthesis is illustrated in **Figure S1**. Mineral wool was soaked in the GO solution and successively squeezed to ensure a complete penetration of the GO solution (Ge et al. 2017). The GO-soaked mineral wool was introduced in a hydrothermal autoclave reactor (Technistro, India), subjected to vacuum for 30 min and placed in an oven for 12 h at 180°C. The resulting graphene-based sponge was then cleaned with MilliQ water, placed in a freezer at -85°C overnight, and freeze-dried. The graphene-based sponges were characterized using scanning electron microscopy (SEM), X-ray photoelectron spectroscopy (XPS), X-ray powder diffraction (XRD), Raman spectroscopy, and other techniques (**Text S1**)

**Electrochemical water treatment experiments**

Experiments were performed in a flow-through, cylindrical reactor made of methacrylate, in one-pass continuous mode. The flow rate was controlled using a digital gear pump (Cole-Parmer) at 5 mL min$^{-1}$ that corresponds to surface area normalized permeate flux of 175 L m$^{-2}$ h$^{-1}$ (LMH) (the projected surface area was 17.34 cm$^2$), and hydraulic residence time (HRT) of the reactor of 3.45 min, unless otherwise specified. The graphene-based sponges of approximately 0.5 cm thickness were fed with current using a 3D stainless steel current collector with spiral geometric design to ensure a homogeneous distribution of current over x-y axis of the sponge and avoid its compression (**Figure S1k**). Wired BRGO and NRGO sponges were employed as anode and cathode, respectively. The anodic stainless steel current collector was stable under the experimental conditions employed in this study. Model organic contaminants (i.e., IPM, DTR, TCL, and DCF) were added to a 10 mM phosphate buffer ($Na_2HPO_4/NaH_2PO_4$, pH 7.2, 1.2 mS cm$^{-1}$) at the initial concentrations of 2 µM. Before applying the current, open circuit (OC) experiments were conducted to verify the loss of target contaminants due to their adsorption onto the graphene-based sponge electrodes. Following the OC, electrochemical degradation experiments were conducted in the chronopotentiometric mode using a BioLogic multi-channel potentiostat/galvanostat VMP-300 and a leak-free Ag/AgCl reference electrode (Harvard Apparatus). The applied anodic currents were 75, 150 and 300 mA (i.e., 43.2, 86.5 and 173 A m$^{-2}$, respectively, calculated using the projected anode surface area). The ohmic drop was calculated from the ohmic internal resistance obtained in the electrochemical impedance spectroscopy (EIS). The EIS experimental data was fitted using the BioLogic EC-lab software (**Text S2**). The impact of the flow direction on the removal of contaminants was studied using the anode-cathode and cathode-anode sequence. To investigate the impact of the flow rate, experiments were performed using the anode-cathode flow direction at 300 mA of the applied anodic current, and at 2.5, 5, and 10 mL min$^{-1}$, resulting in the HRTs of 1.72, 3.45 and 6.9 min, respectively. To investigate the impact of chloride, experiments were performed with 20 mM NaCl added to the 10 mM phosphate buffer. To investigate the formation of ozone, experiments were performed in 10 mM (1.2 mS cm$^{-1}$) and 100 mM (11 mS cm$^{-1}$) phosphate buffer as supporting electrolyte, in both flow directions, at 300 mA of applied anodic current. The formation of OH$^{\bullet}$ was evaluated in the anode-cathode configuration at 75, 150 and 300 mA of anodic current, and using terephthalic acid (TA) as OH$^\bullet$ probe compound (20 mg L$^{-1}$, in 10 mM phosphate buffer, pH 7). TA does not react via direct electrolysis (Jing and Chaplin 2017) and has very low reactivity with ozone (Zang et al. 2009). Preliminary experiments confirmed no adsorption of TA onto the graphene sponge electrodes. Thus, TA was employed as a probe compound for electrochemically generated OH$^\bullet$ ($k_{TA,OH}=4\times10^9$ M$^{-1}$ s$^{-1}$) (Charbouillot et al. 2011). The quasi steady state OH$^\bullet$ concentration, [OH$^\bullet$]$_{SS}$ was determined as a ratio of pseudo-first rate constant of TA decay ($k_{TA}$, s$^{-1}$) and $k_{TA,OH}$ (Moura de Salles Pupo et al. 2020, Nayak and Chaplin 2018). To confirm the electrochemical degradation of an adsorbed contaminant, the employed sponges were extracted with 30 mL methanol, that was evaporated to dryness, reconstituted to a 1 mL sample with methanol/water (25/75, v/v) and analyzed for the presence of target contaminants. All experiments were performed in triplicate and the results were expressed as mean with their standard deviation. To gain insight into the degradation pathways, experiments were performed using IPM as a model compound at high initial concentration (25 µM) to allow the detection of transformation products (TPs) in the full-scan exploration mode. To differentiate the TPs formed at the BRGO anode and NRGO cathode, sampling was conducted after the anode by adding a sampling syringe between the electrodes isolated with the fine polypropylene meshes, and at the exit of the electrochemical reactor. The reactor was operated at 300 mA of applied current, and samples were taken at 10, 20, 30, 40 and 50 bed volumes to ensure that the reactor had reached a steady state. To verify the stability of the graphene-based sponges, the electrochemically treated samples were analyzed for the presence of GO and RGO by measuring the absorbance at wavelengths of 231 nm and 270 nm using UV-Vis diode array spectrophotometer (Agilent Technologies) (Çiplak et al. 2015, Rabchinskii et al. 2016).

**Analytical methods**

Target organic contaminants were analyzed with a 5500 QTRAP hybrid triple quadrupole-linear ion trap mass spectrometer with a turbo Ion Spray source (Applied Biosystems), coupled to a Waters Acquity Ultra-Performance$^{TM}$ liquid chromatograph (Milford). The details of the target analytical method are given in **Text S3** and **Table S1**. Free chlorine and ozone were measured in the presence and absence of 20 mM NaCl, respectively, immediately after sampling and using a diethyl-p-phenylene diamine (DPD) colorimetric method, i.e., Chlorine/Ozone/Chlorine dioxide cuvette tests LCK 310 (Hach Lange Spain Sl). The detection limit for the measurement of free chlorine (HOCl/OCl$^-$) using the DPD method was 0.05 mg L$^{-1}$. Chloride, chlorate and perchlorate were measured by high-pressure ion chromatography (HPIC) using a Dionex ICS-5000 HPIC system. The quantification limits for the measurements of Cl$^-$, ClO$_3^-$ and ClO$_4^-$ were 0.025 mg L$^{-1}$, 0.015 mg L$^{-1}$ and 0.004 mg L$^{-1}$, respectively. Hydrogen peroxide (H$_2$O$_2$) was measured by a spectrophotometric method using 0.01 M copper (II) sulphate solution and 0.1% w/v 2,9 dimethyl -1,10 - phenanthroline (DMP) solution, based on the formation of Cu(DMP)$^+_2$ cation, which shows an absorption maximum at 454 nm (Baga et al. 1988).

## Results and discussion

**Characterization of the graphene-based sponges**

SEM analyses of the synthesized graphene-based sponges confirmed the coating of the mineral wool and presence of the wrinkled graphene sheets on its surface (**Figure 1, Figure S2**). The measured RGO loading of the graphene sponge was 0.038±0.002 g of RGO per g mineral wool (result for the synthesis of 4 sponges). The employed GO solution (i.e., ~0.05 g of GO per g of mineral wool) provided excess GO to ensure complete coating of the mineral wool. The XPS analysis revealed that the hydrothermal reduction increased the C/O atomic ratio from 1.70 for GO, to 3.46 and 4.80 for BRGO and NRGO, respectively (**Figure 1c-f, Figure S3a, Table S2**).

The total N content of the NRGO was 6.4% and was identified as pyridinic-N (2.5%; 398.4eV), pyrrolic-N (3%; 399.8eV), graphitic-N (0.38%; 401.6eV) and azide groups (0.45%, 402,6 eV) (**Table S3**). For the BRGO, the measured content of atomic boron was 1.3%. The sensitivity of the XPS analysis for trace amounts of boron is limited (Wang et al. 2013), in particular for BRGO coating on a mineral wool substrate. Although the boron content was low, it was only present when boric acid was used in the synthesis process. In addition, BRGO and undoped RGO also contained around 1% of N atom, mainly pyrrolic-N and pyridinic-N, originating from the commercial GO solution employed (**Table S2 and S3**). Thus, BRGO can be considered a co-doped graphene-based sponge, with B and N atomic functionalization in its structure.

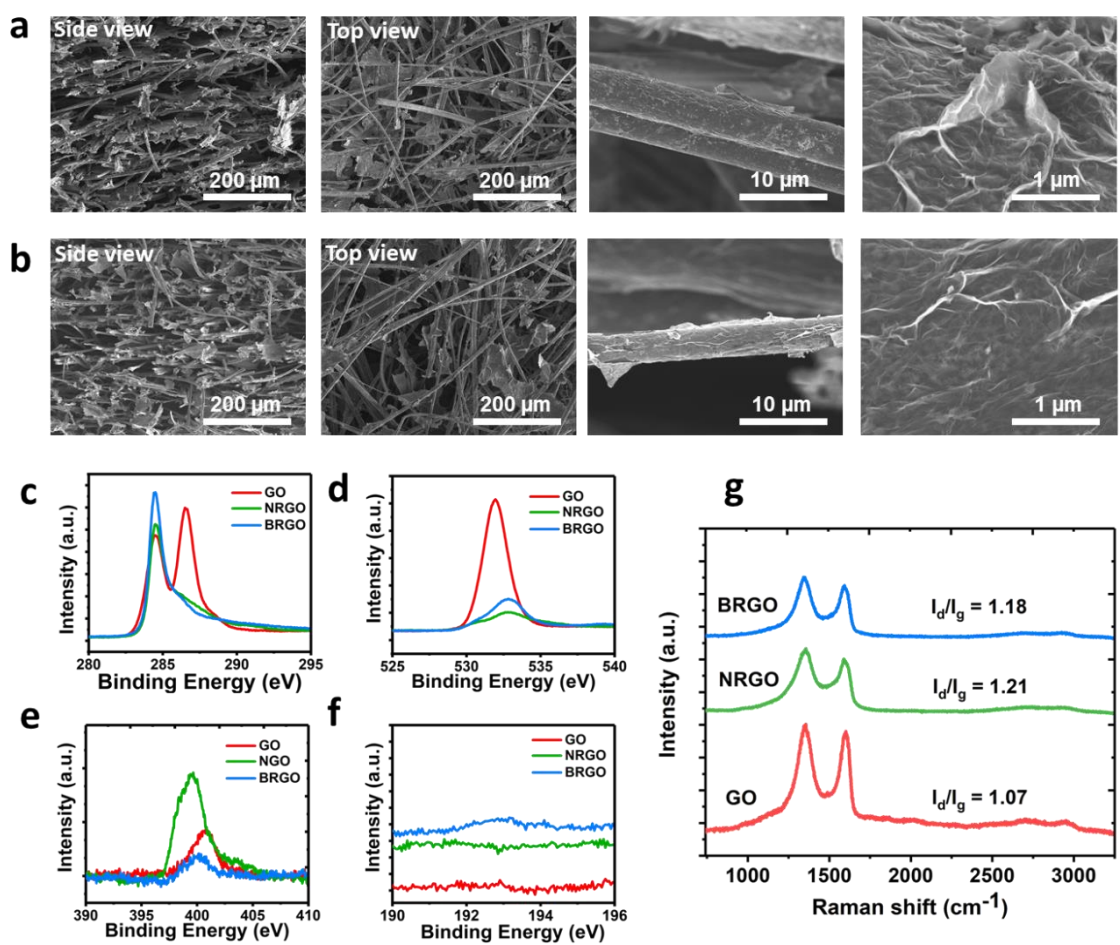

**Figure 1.** SEM images of **a)** BRGO, and **b)** NRGO. **c)** C1s, **d)** O1s, **e)** N1s and **f)** B1s XPS spectra of GO, BRGO and NRGO. **g)** Raman Spectroscopy profiles of GO, BRGO and NRGO.

The level of defects was calculated by measuring the intensity ratio of D peak at 1347 cm$^{-1}$ ($I_d$) and a G peak at 1581 cm$^{-1}$ ($I_g$) obtained using Raman spectroscopy (**Figure 1g**). The higher the $I_d/I_g$ ratio is, the higher is the content of the graphene defects. The measured $I_d/I_g$ ratios were 1.07 for GO, 1.18 for BRGO and 1.21 for NRGO. Higher $I_d/I_g$ ratio for the BRGO and the NRGO is a result of the reduction of GO and indicates a defect-rich structure, which provides additional (electro)catalytic active sites (Wang et al. 2016). The XRD analysis showed a decrease in the interlayer spacing from 8.1 Å for GO to ≈ 3.5 Å for BRGO and NRGO, due to the removal of the oxygen functional groups from the basal plane (**Figure S3b**). The BET surface areas were 0.81 m$^2$ g$^{-1}$ for the mineral wool template, 1.39 m$^2$ g$^{-1}$ for the BRGO, and 1.44 m$^2$ g$^{-1}$ for NRGO (**Figure S3c**). The increase in the specific surface area of the graphene-based sponges relative to the mineral wool template can be assigned to the wrinkled structure of the RGO coating. These values are significantly lower compared to the literature values for other 3D graphene networks (e.g., graphene-based hydrogels and aerogels) due to the use of the supporting template (i.e., mineral wool) (Cong et al. 2012). The interaction of a water droplet with the BRGO and NRGO sponges is represented in **Figures 2a** and **2b**, respectively. The contact angles determined for BRGO and NRGO were 139.67º±4.50º and 137.67º±7.50º, respectively, demonstrating a high degree of hydrophobicity of the sponges, similar to the previously reported values (Ge et al. 2017). The through-plane resistances for BRGO and NRGO were 16.05±5.82 kΩ and 55.33±13.65 kΩ, whereas the in-plane resistances were of 2.26±1.48 kΩ and 7.48±4.01 kΩ, respectively (**Figure 2c**). Higher through-plane than in-plane resistance is a consequence of the orientation of the mineral wool fibers, with less fibers connecting the parallel sheets of the mineral wool and thus less graphene deposited between these sheets (**Figure 1g**). This can be disadvantageous in the case of planar current collectors because the distribution of current in the y-axis is limited and was overcome in this study by using 3D current collectors (**Figure S1k**). Also, graphene-based sponges need to be carefully handled when dry (e.g., after the freeze drying step) to avoid macroscopic defects and thus increase in resistance. Once the sponges are wet, they become flexible and can be easily handled. The three-fold higher resistance of the NRGO in comparison with the BRGO can be explained by the higher content of the N-dopant of the former, and thus higher disruption of the graphene structure by the pyridinic-N and pyrrolic-N (Sun et al. 2012). The conductivity of the graphene-based sponges is illustrated by the illumination of a LED lamp (inset **Figure 2c**).

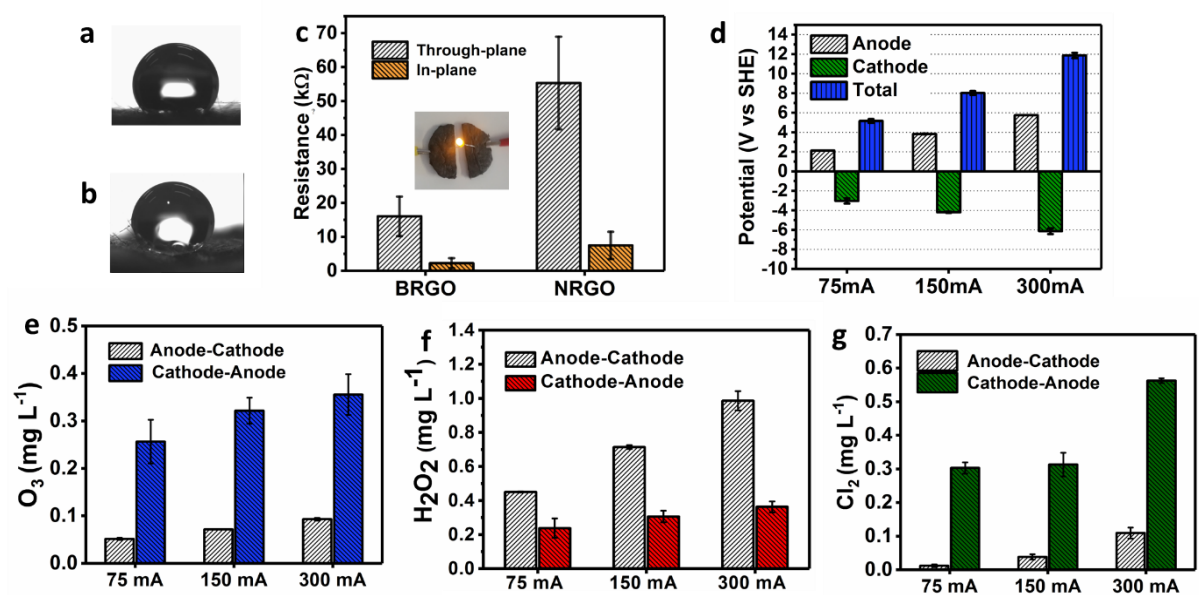

**Figure 2.** Contact angle measurement of **a)** BRGO and **b)** NRGO, **c)** through-plane and in-plane resistance of BRGO and NRGO sponges (inset: lightning a LED using graphene sponge as a connector), **d)** recorded anode, cathode and total cell potential at different applied currents for BRGO-NRGO electrochemical reactor; generation of oxidant species in BRGO-NRGO electrochemical filter at different applied currents and flow directions of **e)** $O_3$ and **f)** $H_2O_2$ formed in 10 mM phosphate buffer, and **g)** free chlorine formed in the presence of 20 mM NaCl.

**Electrochemical performance of graphene-based sponge electrodes**

Application of 75 mA, 150 mA and 300 mA of anodic current did not result in any visual or electrical degradation of graphene-based sponges and yielded stable anodic potentials of 2.1±0.06, 3.8±0.1 and 5.7±0.02 V/SHE (ohmic-drop corrected anodic potentials of 1.6, 2.8 and 3.7 V/SHE, based on the EIS measurements), respectively (**Figure 2d**, **Table S4**). The pH remained constant at pH 7 in all experiments suggesting that the anodic production of $H^+$ was likely compensated by the cathodic production of $OH^-$, eliminating the need for the pH correction of the treated water.

The placement of the electrodes will have a determining impact on the electrogeneration of oxidant species. For example, $O_3$ and chlorine generated at the anode can be easily reduced at the subsequent cathode in the anode-cathode flow direction. The electrogeneration of $H_2O_2$ may be enhanced in the anode-cathode configuration due to the cathodic reduction of the anodically generated $O_2$ - although the accumulation of bubbles may hinder the $O_2$ mass transfer (Zhou et al. 2018). Thus, we investigated the formation of the above mentioned oxidants in the anode-cathode and cathode-anode reactor configurations at different applied anodic currents, and at 5 mL min$^{-1}$ flowrate (HRT=3.45 min). As shown in **Figure 2e**, at 75 mA the measured concentration of $O_3$ was 0.05±0.002 mg L$^{-1}$ in the anode-cathode direction. When the placement of the electrodes was reversed, the $O_3$ concentration was 0.3±0.05 mg L$^{-1}$ (current efficiency 0.13%, **Text S4**). Further increase in the applied current to 300 mA enhanced the generation of $O_3$ to 0.09±0.003 mg L$^{-1}$ in the anode-cathode flow direction, and 0.4±0.04 mg L$^{-1}$ for the cathode-anode configuration. These results confirmed: *i)* the electrogeneration of $O_3$ at the BRGO anode, proportional to the applied anodic current, and *ii)* the reduction/decomposition of the formed $O_3$ at subsequent cathode in the anode-cathode flow direction. Recent study reported the electrocatalytic production of $O_3$ at the N and B co-doped mesoporous carbon electrode, with 3.5 mg L$^{-1}$ of $O_3$ formed at 700 A m$^{-2}$ of anodic current density applied in saturated $K_2SO_4$ (~726 mM), i.e., in a highly conductive supporting electrolyte (>85 mS cm$^{-1}$) (Zhang et al. 2020). The BRGO anode contains 1.3% of boron doping, 0.5% of pyrrolic-N in its structure, and an $I_d/I_g$ ratio of 1.18 demonstrating the presence of defects in its structure (**Tables S2, S3**). These sites may have a similar synergy in the anodic generation of ozone as the one reported by Zhang et al (Zhang et al. 2020). To investigate the impact of the electrolyte conductivity on the formation of $O_3$, experiments were also performed in the 100 mM phosphate buffer. The concentration of the formed $O_3$ in cathode-anode configuration at 300 mA was 0.7 mg $L^{-1}$, i.e., two-fold higher compared to the low-conductivity 10 mM phosphate buffer (0.3 mg $L^{-1}$) (**Figure S4**). Thus, the developed B-doped graphene-based sponges used with the high-conductivity supporting electrolytes may find potential application for the *in situ* electrochemical ozone production, although this was out of scope of the current study.

In terms of $H_2O_2$ formation (Figure 2f), anode-cathode configuration yielded higher $H_2O_2$ concentrations (e.g., 0.45±0.003 mg $L^{-1}$ at 75 mA) than cathode-anode configuration (e.g., 0.24±0.06 mg $L^{-1}$ at 75 mA). In the cathode-anode flow direction, the cathodically formed $H_2O_2$ is decreased due to its reaction with the anodically formed $O_3$. Also, as previously reported (Zhou et al. 2018), anodic generation of $O_2$ and its posterior reduction at the cathode also yields higher $H_2O_2$ concentration. Like in the case of ozone, higher $H_2O_2$ concentrations were measured at higher applied currents, reaching 1±0.1 mg $L^{-1}$ in anode-cathode configuration at 300 mA. Yet, these concentrations are likely only presenting the residual amount of $H_2O_2$ due to its reaction with the anodically formed ozone, as well as the presence of the pyrrolic-N and pyridinic-N in the NRGO cathode that activate the decomposition of the formed $H_2O_2$ to $OH^{\bullet}$ (Yang et al. 2018, Yang et al. 2019). Based on the obtained results from the experiments with TA as $OH^{\bullet}$ probe, the estimated steady state concentration of $OH^{\bullet}$ in the BRGO(A)-NRGO(C) system at 300 mA of anodic current was 9.7 x $10^{-14}$ M.

A major limitation of all commercial anode materials is rapid production of chlorine in the presence of chloride, and formation of persistent and toxic chlorinated byproducts (Chaplin 2019, Radjenovic and Sedlak 2015). The most often used commercial anode materials in the electrochemical water treatment (i.e., BDD, $Ti_4O_7$ and $SnO_2$-based mixed metal oxide electrodes) are also very efficient in generating chlorate, perchlorate and chlorine, which reacts rapidly with the organic matter to form organochlorine byproducts (Bagastyo et al. 2011, Radjenovic et al. 2011, Wang et al. 2020). The formation of perchlorate may be limited for lower concentrations of chloride (i.e., up to 5 mM) at the flow-through $Ti_4O_7$ anodes, although persistent polychlorinated organic byproducts were still present in the treated effluent (Lin et al. 2020). In our experiments conducted without the added contaminants, the highest concentration of free chlorine measured at the applied current of 300 mA, in the cathode-anode direction and in the presence of 20 mM NaCl was only 0.56±0.01 mg $L^{-1}$ (**Figure 2g**). Such high initial concentration of $Cl^-$ was intentionally selected to build up chlorine that may be formed at the anode and is equivalent to the salinity of reverse osmosis brines from water reclamation plants. In the anode-cathode configuration at the same applied current, free chlorine concentration was somewhat lower, 0.1±0.02 mg $L^{-1}$, due to the cathodic reduction of the formed chlorine. It should be noted here that the DPD colorimetric method used for chlorine measurement cannot differentiate between the contributions of individual oxidants. Thus, the obtained values may overestimate the actual chlorine concentrations due to the presence of ozone, which was only determined in the chloride-free supporting electrolyte but can also be formed in the presence of NaCl. The maximum calculated current efficiency for chlorine formation in the cathode-anode direction was only 0.04% (**Text S4**). Moreover, we did not detect observe $ClO_3^-$ and $ClO_4^-$ formation, further demonstrating that the electrooxidation of chloride was very limited. Additionally, cyclic voltammetry was performed with the BRGO anode and did not show an increase in current when switching from 10 mM phosphate buffer to 8 mM NaCl solution (**Figure S5**), indicating poor electrocatalytic activity of our graphene-based sponge electrode towards chloride oxidation. The existing knowledge on the interactions of halide ions and graphene-based materials is limited to density functional theory (DFT) calculations and it is unclear why chloride ions are not oxidized at the RGO-coated anode (Farajpour et al. 2016, Zhu and Yang 2016). Chloride ion is expected to be weakly adsorbed on the graphene surface (Shi et al. 2012, Zhu and Yang 2016). DFT modelling of the interaction of halide ions with graphene in the presence of electric field reported the stabilization of $Cl^-$ anion on the edge area of the graphene flake via dipole-dipole interactions (Cole et al. 2011, Farajpour et al. 2016). Also, the simulation study by Cole et al (Cole et al. 2011) indicated significant interaction between $H_3O^+$ and $Cl^-$ ions at the graphene (or graphene oxide)/water interface and stabilization of their ion-ion pairs. Since hydronium ions are produced in water electrolysis at the BRGO anode ($6H_2O \rightarrow O_2(g) + 4H_3O^+(aq) + 4e^-$, $E° = +1.23$ V), their interaction with $Cl^-$ ions may lead to their stabilization at the graphene sponge anode surface, thus limiting the electrooxidation of $Cl^-$. Although the exact mechanism of $Cl^-$ interactions with the developed graphene sponge electrodes remains unclear, experimental evidence obtained in this study demonstrates that RGO-based sponges and similar graphene-based electrodes may overcome the major limitation of existing anode materials – formation of chlorinated byproducts in the presence of chloride.

The UV-vis analyses of the samples treated at 300 mA of applied anodic current were analyzed using UV-vis spectroscopy, as GO and RGO show absorption peaks at 230-232 nm and 269-270 nm, respectively (Çiplak et al. 2015, Rabchinskii et al. 2016). No RGO release could be detected at the highest applied anodic current (i.e., 300 mA), with the detection limit for RGO and GO of 0.018 mg $L^{-1}$ (**Figure S6**). Thus, there was no loss of the RGO coating under the conditions of the experiment reported in this study. Mineral wool is mainly comprised of $SiO_2$ (Väntsi and Kärki 2014), and the stability of the produced materials towards electrochemical polarization is likely due to the covalent bonding between the graphene-based sheets and $SiO_2$ (Hintze et al. 2016, Ramezanzadeh et al. 2016). Furthermore, DFT modelling studies predicted the formation of C-O and C-Si covalent bonds and thus a strong interfacial adhesion between graphene and $SiO_2$ (Shemella and Nayak 2009).

**Electrochemical removal of persistent organic contaminants**

The removal of trace organic contaminants obtained in the anode-cathode and cathode-anode flow directions in the OC and at varying anodic currents is presented in **Figures 3a and b,** respectively. The values presented are mean values, whereas the obtained mean relative standard deviations are lower than 5%. The effluent concentrations of TCL and DCF were decreased for ~80% and 20%, respectively, due to their adsorption at the graphene-based sponges. Almost complete adsorption of TCL under OC conditions can be explained by its high hydrophobicity (logD=5.13, pH 7.4, **Table S5**). Hydrophobic effect and π-π interactions have a determining effect on the adsorption of hydrophobic aromatic organics to carbon nanomaterials (Yang and Xing 2010, Zhao and Zhu 2020). Apart from TCL, 10-20% removal due to adsorption was observed for DCF (logD = 1.37, pH 7.4, **Table S5**). Recent study suggested that the adsorption of DCF on graphene materials is mediated by the hydrogen bonds with the functional groups of RGO, and electrostatic interactions (Jauris et al. 2016). The effluent concentrations of IPM and DTR were equal to their feed concentrations in the OC, indicating that these contaminants were not adsorbed onto the graphene-based sponges, in agreement with their very low LogD values (logD = -2.12, pH 7.4, **Table S5**).

When current was applied, the anode-cathode configuration resulted in the higher removal for all target contaminants compared with the cathode-anode direction (**Figure 3 a, b**). At the highest applied current of 300 mA (5 mL min$^{-1}$, i.e., HRT=3.45 min), the obtained removal efficiencies of IPM, DCF, TCL, and DTR in the anode-cathode configuration were 91±3%, 88±1%, 99±0.01%, and 84±2% (**Figure 3a**) and 54.6±4%, 39.2±2%, 74.6±2%, and 37.9±6% in the cathode-anode configuration (**Figure 3b**), respectively. The cathode-anode configuration was also performing worse in terms of reaching the steady state within the selected sampling points for each current (i.e., up to 30 bed volumes). Better performance of the anode-cathode flow direction can be explained by several factors: *i)* enhanced production of $H_2O_2$ via cathodic reduction of $O_2$ produced at the upstream anode; and *ii)* enhanced activation of $O_3$ generated at the anode by $H_2O_2$ produced at the cathode (Staehelin and Hoigne 1985). The amount of generated ozone at BRGO anode was likely similar in both anode-cathode and cathode-anode configurations, but it could only be measured when $O_3$ is not quenched by the second electrode (i.e., cathode), as shown in **Figure 2e**.

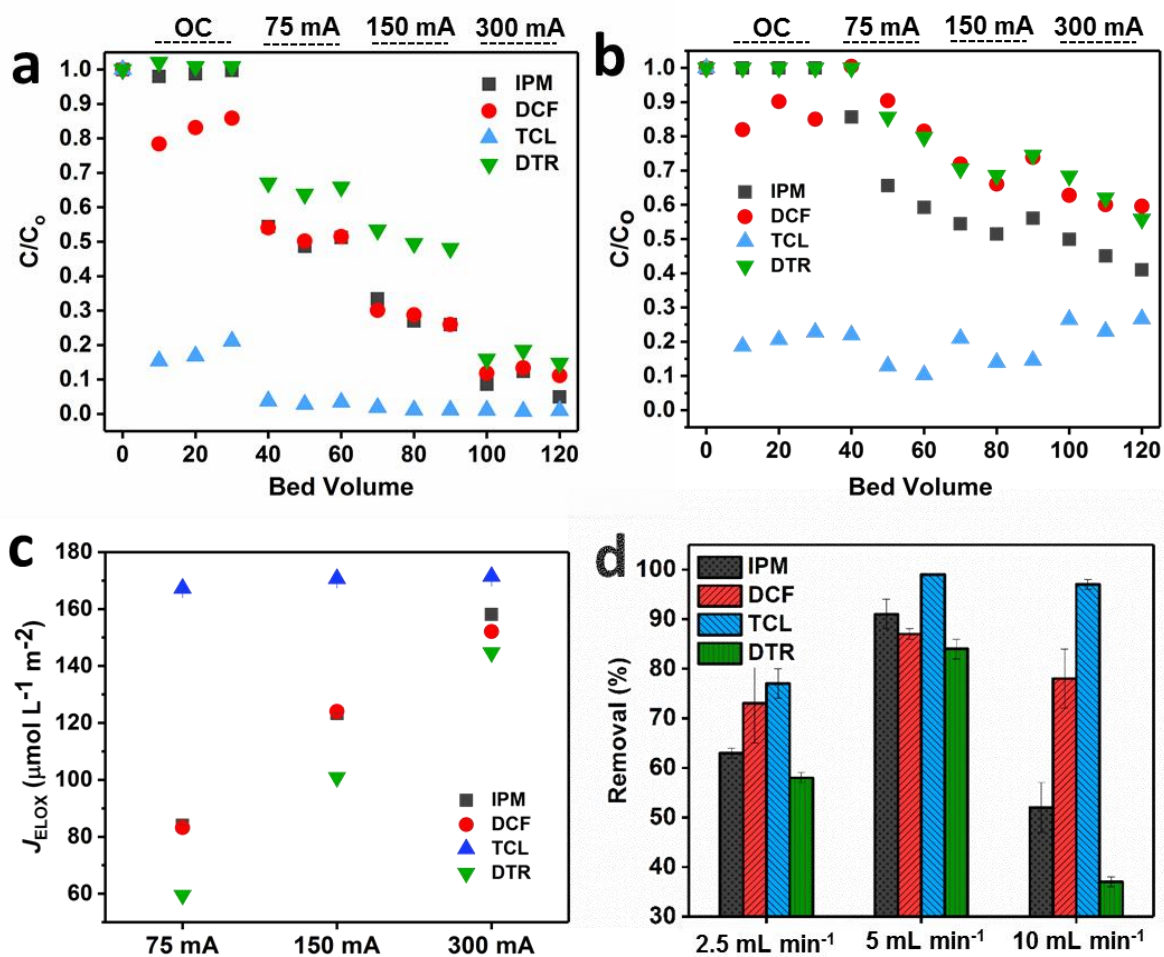

**Figure 3.** Electrochemical removal of persistent organic contaminants using graphene-sponge electrodes, with effluent concentrations (C) normalized to the feed concentrations ($C_0$) of target contaminants in the open circuit (OC) and at different anodic currents and constant flow rate of 5 mL min$^{-1}$ in: **a)** anode-cathode and **b)** cathode-anode directions; **c)** electrooxidation fluxes ($J_{ELOX}$) of target contaminants obtained in the anode-cathode direction at different anodic currents and constant flow rate of 5 mL min$^{-1}$; **d)** removal % vs flow rate at 300 mA of applied current in the anode-cathode direction.

In the anode-cathode configuration, the removal of TCL (i.e., 99%) was somewhat improved relative to the OC (i.e., 82%). Pronounced adsorption of TCL onto the graphene-based sponge electrodes made it difficult to identify its electrochemical degradation. Nevertheless, TCL likely reacted with the electrogenerated $O_3$ and $OH^•$, given its reported bimolecular rate constants with these oxidants (**Table S6**). In the case of IPM, DTR, and DCF, increase in current from 75 to 150 and 300 mA resulted in the stepwise increase in their removal efficiencies (**Figure 3a**). Moreover, the calculated electrooxidation fluxes of DTR, IPM and DCF were linearly correlated with the applied anodic current (**Figure 3c**). For example, the removal of DTR, highly polar and recalcitrant contaminant was increased from 34±1% (at 75 mA) to 58±2% (at 150 mA) and 84±2% (at 300 mA). Given that this compound is persistent towards $O_3$ and difficult to oxidize by the $OH^•$ radicals (**Table S6**), it is likely that direct electrolysis played a significant role in its electrochemical degradation. In addition, DTR is negatively charged at bulk pH 7, therefore it may have initially electrosorbed to the BRGO anode. IPM also showed no adsorption affinity towards the graphene-based sponges and its removal was observed only after the application of current, reaching 48±2%, 71±4% and 91±3% at 75, 150 and 300 mA, respectively. IPM was present in its uncharged form at pH 7 (pKa=11.4, **Table S5**) and although electrosorption of the uncharged species is rather low, it still may occur due to their polarizability and dipole formation under the external electric field (Niu and Conway 2002, Thamilselvan et al. 2018). Like DTR, this iodinated contrast agent is recalcitrant to $O_3$ but is more reactive with $OH^•$ (**Table S6**) and thus was likely removed by the combination of direct electron transfer and oxidation via $OH^•$. Increase in the anodic current was also linearly correlated with the removal of DCF (**Figure 3a, b**), which was likely oxidized via the electrogenerated $O_3$ (6.8 x $10^5$ $M^{-1}$ $s^{-1}$) (Sein et al. 2008) and $OH^•$ ($9.3 \times 10^9$ $M^{-1}$ $s^{-1}$) (Yu et al. 2013) (**Table S6**), and possibly via direct electron transfer. In summary, the increase in the observed removal efficiencies of the contaminants with low adsorption affinity (DCF) or no adsorption affinity (IPM, DTR) with the increase in the applied anodic current (**Figure 3c**) suggests that these contaminants undergo electrocatalytic degradation. This was reinforced by the measurements of the last OC run conducted after several electrochemical runs at different current densities, as there was no increase in the effluent concentrations above the feed concentrations after turning off the current (**Figure S6**). Thus, even if initially electrosorbed, the model contaminants were further degraded.

The calculated electrooxidation fluxes of the trace contaminants in the anode-cathode configuration at 300 mA of applied current were 171.4±0.001, 152.1±0.01, 158.08±0.04 and 144.7±0.02 µmol L$^{-1}$ m$^{-2}$ for TCL, DCF, IPM and DTR, respectively (**Figure 3c**). The obtained electrooxidation fluxes are two orders of magnitude lower compared to previously reported values for the removal of organic contaminants using a CNT-filter (Liu et al. 2015). Yet, electrochemical performance of the CNT-filter was studied using three to four orders of magnitude higher feed concentrations of the contaminants and high conductivity electrolyte, thus making it difficult to compare the data with our study. The electric energy per order ($E_{EO}$) for the removal of target contaminants at 300 mA was calculated using the total cell potential, and it was 5.8±0.3 kWh m$^{-3}$ for TCL, 11.2±0.3 kWh m$^{-3}$ for IPM, 12.9±0.4 kWh m$^{-3}$ for DCF, and 15.1±0.12 kWh m$^{-3}$ for DTR. High $E_{EO}$ values are a consequence of high total cell potentials obtained at 300 mA (**Table S4**) due to the accumulation of the produced $O_2$ and $H_2$ bubbles and increased ohmic resistance. Although strong interaction of TCL with the graphene sponges makes the elucidation of its electrochemical degradation difficult, this behavior is beneficial from the application point of view as it decreases the mass transfer limitations and lowers the energy consumption. Furthermore, extraction of the employed sponges with methanol did not show any trace concentrations of TCL or other trace contaminants, although in the case of IPM, DTR and DCF their absence in the methanol extract is expected as these contaminants do not adsorb onto the graphene-based sponges. Given that the preliminary tests showed near complete recovery (>95%) of TCL in the extraction with methanol, this result further confirms that the adsorbed TCL underwent electrochemical degradation/transformation.

Increasing the flow rate from 5 to 10 mL min$^{-1}$ had a detrimental impact on the removal of trace contaminants with low adsorption affinity, with 38±1%, 52±5% and 78±6% removal obtained for DTR, IPM and DCF, respectively, whereas the removal of hydrophobic TCL was not significantly affected (**Figure 3d**). These results indicate that the removal of organic contaminants with poor adsorption affinity towards the graphene-based sponges was limited by their mass transfer and diffusion to the electrode surface. The convection-enhanced mass transfer of pollutants is typically achieved at the CNT-based electrodes and porous $Ti_4O_7$ electrodes with very small pore diameters (Trellu et al. 2018). In this study, the porosity of the graphene-based sponge is determined by the employed mineral wool template and resulted in 10-50 µm average pore size. Another reason for the negative impact of the higher flowrate on the contaminant removal may also be a favored oxygen evolution reaction due to the sweeping of the gas bubbles from the electrode surface (Trellu et al. 2018). When the flow rate was decreased to 2.5 mL min$^{-1}$ (i.e, HRT=6.9 min), it also led to somewhat worsened removal of trace contaminants compared to the experiments conducted at 5 mL min$^{-1}$. This was due to the poor evacuation of the electrogenerated $H_2$ and $O_2$ bubbles and their accumulation within the graphene-based sponges, which led to the poor wetting of the electrodes and thus decreased their interaction with the contaminants.

To demonstrate the beneficial effect of the atomic doping of the graphene-based sponges, experiments were also performed with the undoped RGO electrodes, i.e., without the addition of boric acid/urea in the synthesis procedure. The characterization of the undoped RGO sponge is summarized in **Text S5**, and **Figures S3** and **S8**. The removal of IPM, DTR and DCF was significantly worsened in the BRGO anode/RGO cathode and the RGO anode/NRGO cathode systems compared with the BRGO anode/NRGO cathode set-up (**Figure S9**). The B- and N-doping did not lead to significant changes in the charge transfer resistance as determined by the EIS. Nevertheless, at all current densities, the lowest total cell potentials in the anode-cathode configuration were recorded for the best-performing BRGO(A)-NRGO(C) system (**Table S4**). The worsened removal of target pollutants in the experiments conducted with the RGO electrodes is in line with the expected enhanced production and activation of $H_2O_2$ at the NRGO cathode (Su et al. 2019, Yang et al. 2018, Yang et al. 2019), and formation of $O_3$ at the BRGO anode (Zhang et al. 2020). In addition, pyridinic and pyrrolic-N as well as B-atoms are p-type dopants and create electron-deficient sites (Schiros et al. 2012, Wang et al. 2013). This redistribution of the electron density impacts the interaction of the contaminants with the graphene-based sponge electrodes in multiple ways (e.g., cation-π interactions, π-π interactions, etc) and, in the case of the target contaminants employed in this study, leads to their enhanced removal.

**Elucidation of the transformation pathway of iopromide**

Comparison of the full-scan mode of analysis of the initial IPM solution and samples taken after the anode revealed several TPs formed at the BRGO anode (**Figure S11a, b**). Three TPs could be identified in the sample taken after the BRGO anode, with the molecular ions $[M+H]^+$ 414.2 (TP414), $[M+H]^+$ 400.2 (TP400) and $[M+H]^+$ 453.1 (TP453) (**Figure S12a**). TP414 displayed the CID $MS^2$ spectrum identical to the previously reported spectra for a completely de-iodinated IPM (**Figure S13a**).[51,52] TP400 was identified as an O-demethylation product of TP414 (**Figure S13b**), whereas the formation of TP453 was assigned to the loss of two iodide substituents and metoxyacetyl amine chain from the IPM molecule (**Figure S13c**). The proposed structures of the identified TPs are illustrated in **Figure 4**. In addition, three more molecular ions, i.e., $[M+H]^+$ 393.2 (TP393), $[M+H]^+$ 471.8 (TP471) and $[M+H]^+$ 627.2 (TP627) were isolated from the full-scan spectra of the samples taken after the BRGO anode. However, the elucidation of their exact structures could not be performed based on the CID MS$^2$ spectra (**Figure S14**).

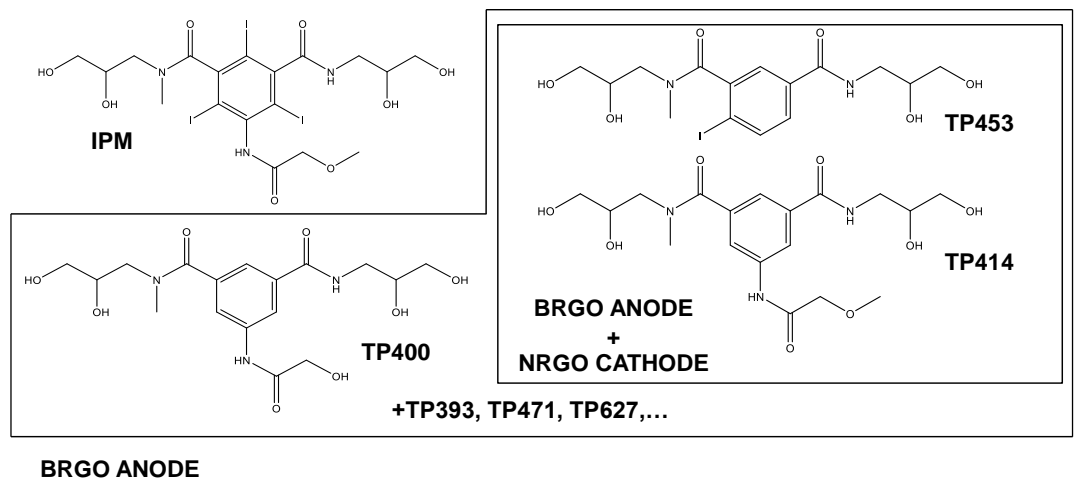

**Figure 4** Transformation products (TPs) of IPM formed after passing through BRGO anode, and BRGO anode + NRGO cathode.

Besides these TPs, full scan analyses indicated other TPs formed at the BRGO anode (**Figure S11b**). Yet, their molecular ions could not be isolated with certainty. The identification of these and other TPs requires the use of high-resolution mass spectrometry and will be the subject of our future studies. Very few TPs could be observed in the total ion chromatogram (TIC) of the final treated sample (**Figure S11c**). For the final effluent samples after the NRGO cathode, the only two TPs encountered were TP414 and TP453 (**Figure S12**).

The cleavage of all three iodide substituents at the aromatic ring is typically achieved in the electroreduction of IPM (Lütke Eversloh et al. 2014, Mu et al. 2010). Here, TP414 was detected in the sample taken after the BRGO anode and with significantly higher signal intensity compared with the sample taken at the reactor exit (**Figure S12**). Oxidative formation of TP414 has not been previously observed. In the electrooxidation of IPM at BDD anodes, the 2,4,6-triiodinated aromatic ring remained intact in most of the identified TPs (Lütke Eversloh et al. 2014, Radjenovic and Petrovic 2016). Although the oxidative degradation of IPM may lead to a near-complete release of iodide (Radjenovic and Petrovic 2016), the initial oxidative cleavages are localized at the alkyl side chains of IPM, and the scission of all three C-I bonds takes place only after the steric hindrance at the aromatic ring has been reduced. Preferential anodic cleavage of all three C-I bonds at the aromatic ring over scissions at the alkyl side chains indicate the interaction of the neutral IPM molecule with the graphene-based sponge anode via mechanisms other than the OH$^\bullet$-induced degradation, typically observed in electrooxidation and other oxidation processes. Similar to other aromatic pollutants, IPM can interact with graphene via π- π interactions (Björk et al. 2010), leading to a temporary "docking" of the aromatic ring of IPM at the graphene-based sponge surface and thus facilitated oxidative cleavage of the iodide substituents. Given that IPM was not removed at all via adsorption in the OC experiments, interaction of IPM with the BRGO anode can be explained by the formation of the IPM dipole under the external electric field, and its electrosorption (Niu and Conway 2002, Thamilselvan et al. 2018).

The formation of TP400 indicated the O-demethylation of TP414. O-demethylation was previously observed in the electrooxidation pathway of IPM at BDD anodes (Radjenovic and Petrovic 2016) and is considered to proceed via H-abstraction by OH$^\bullet$ or other oxidative radical species (Chahboune et al. 2015). In addition, the cleavage of the C-N bond and the loss of metoxyacetyl amine chain evidenced in the formation of TP453 is another indication of the OH$^\bullet$-induced oxidation of IPM. N-dealkylation and subsequent loss of the amine group was previously observed in the electrooxidation of IPM (Radjenovic and Petrovic 2016) and OH$^\bullet$-based oxidation of other contaminants having an N-alkyl aromatic substituent (Khan et al. 2014). Thus, OH$^\bullet$ radicals participated in the oxidative degradation of IPM, although the origin of the formed OH$^\bullet$ could not be determined. In the case of BDD and other electrode materials (e.g., $Ti_4O_7$, $Ti/SnO_2$-Sb), electrochemical generation of OH$^\bullet$ at the hydrophilic anode surface is considered to proceed via the direct electrolysis of water (Comninellis et al. 2010). In the case of the highly hydrophobic graphene-based sponge anode, it is more likely that the OH$^\bullet$ are formed by the decomposition of $O_3$ formed at the BRGO anode (**Figure 2e**). The absence of various TPs identified at the BRGO anode in the final reactor effluent indicates their further oxidation at the NRGO cathode via the OH$^\bullet$ radicals formed by the cathodic formation and activation of $H_2O_2$, as explained previously and demonstrated in other studies (Su et al. 2019, Yang et al. 2018, Yang et al. 2019). Further studies are needed to evaluate the toxicity of the TPs of IPM, and whether they would be present in the final treated effluent at low, environmentally relevant concentrations of IPM. The postulated mechanism of the electrochemical degradation of target pollutants is based on the: *i*) *in situ* generation of $O_3$ and possibly other reactive oxygen species (e.g., OH$^\bullet$ formed by $O_3$ decomposition) at the BRGO anode, *ii*) *in situ* generation of $H_2O_2$ at the cathode and its activation to OH$^\bullet$ via the N-active sites at the NRGO cathode; and via the reaction of $H_2O_2$ with the anodically formed $O_3$, and *iii*) direct electron transfer between the target pollutants and the graphene-based sponge electrodes through the functional groups and defects in the graphene structure (**Figure S1** and **Figure S10**).

## Conclusions

In this study, we synthesized atomically doped graphene-based sponge electrodes and employed them for electrochemical removal of a set of model persistent organic contaminants from low conductivity supporting electrolyte. Electrochemical degradation proceeds via anodically generated $O_3$, cathodically generated and activated $H_2O_2$ and likely via direct electrolysis. The exact contributions of each mechanism to the electrochemical degradation of contaminants on the graphene sponges needs to be further studied, as it will also be greatly impacted not only by their reactivity with specific oxidants (e.g., $O_3$, OH$^\bullet$) but also by the extent and nature of adsorption/electrosorption of the contaminants onto the sponge.

The most remarkable result of the present study is the very low electrocatalytic activity of the graphene-based sponge anode for $Cl^-$ oxidation, as there was no chlorate and perchlorate formation, and current efficiency for chlorine production in 20 mM NaCl was only 0.04%. Thus, electrochemical system based on graphene-based sponge electrodes opens the possibility to overcome a major limitation of all currently available electrode materials – formation of organic and inorganic chlorinated byproducts in the electrochemical water treatment. Moreover, graphene-based sponge electrodes developed in this study were produced using a simple, scalable, and low-cost hydrothermal self-assembly method. Currently, high purity graphene oxide is sold at prices of €1,200-1,500/kg GO. The employed loading of GO was ~0.05 g of GO per g of mineral wool (obtained loading: 0.038 g of RGO per g of mineral wool), thus indicating an estimated material cost of €0.08 per gram of mineral wool, calculated for the GO loading employed in excess. The cost of the mineral wool was not considered as it is negligible (less than 50 ¢/kg mineral wool). Thus, this would result in a cost of €46 per $m^2$ of the projected electrode surface area, which makes graphene-based sponges developed in this study very cost-competitive compared with the state-of-the-art BDD anodes with an approximate cost of €6,000 per $m^2$. Given their inherent flexibility and compressibility, they can be applied in different reactor geometries (e.g., Swiss roll, column-type reactors, and others). The introduction of dopants (e.g., B, N and other atoms, two-dimensional materials) into the graphene-based structure can be used to functionalize the surface and tailor the electrocatalytic activity of the graphene-based sponge electrodes for the removal of specific groups of organic, inorganic, and microbial contaminants. Finally, application of graphene-based materials for electrochemical water treatment may uncover synergies with other fields such as energy harvesting and storage and play a significant role in the future development of independent, self-powered water treatment systems.

**Supplementary Material**

Graphene-based sponge materials characterization (SEM, RAMAN, XRD, BET surface area contact angle, CV). Description of the analytical method for target contaminants, and calculations of figures of merit. Ozone formation and electrochemical degradation of pollutants. Chemical structures and physico-chemical properties of target contaminants. The optimized compound-dependent MS-parameters. $MS^2$ mass spectra, total and extracted ion chromatograms of the TPs of IPM.

**Acknowledgments**

This work has been funded by ERC Starting Grant project ELECTRON4WATER, project number 714177. ICRA researchers thank funding from CERCA program. We would like to thank the anonymous reviewers of this manuscript as they helped improve it in a significant manner.

Luis Baptista-Pires[†,‡], Giannis-Florjan Norra[†,‡], Jelena Radjenovic [†,§,*]

[†]*Catalan Institute for Water Research (ICRA), Emili Grahit 101, 17003 Girona, Spain*

[‡]*University of Girona, Girona, Spain*

[§]*Catalan Institution for Research and Advanced Studies (ICREA), Passeig Lluís Companys 23, 08010 Barcelona, Spain*

*\* Corresponding author:*

*Jelena Radjenovic, Catalan Institute for Water Research (ICRA), Emili Grahit 101, 17003 Girona, Spain*

Phone: + 34 972 18 33 80; Fax: +34 972 18 32 48; E-mail: jradjenovic@icra.cat

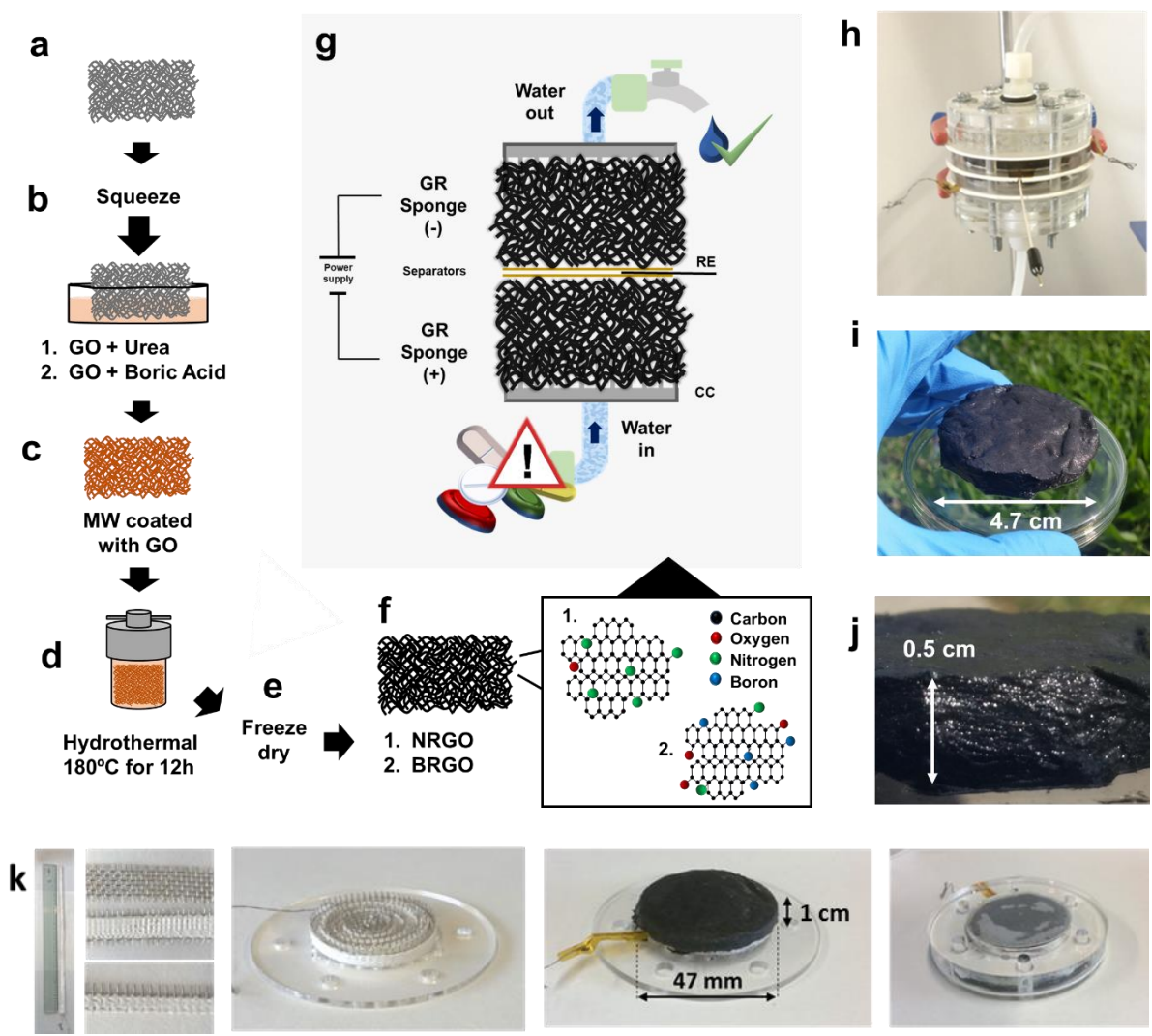

**Figure S1.** Schematic illustrations of: **a-f)** methodology for the production of graphene-based sponge based on hydrothermal reaction of graphene oxide (GO) dispersion in the presence of mineral wool (MW), **g)** electrochemical flow-through treatment system; and images of **h)** flow-through reactor, and **i,j)** graphene sponge; **k)** methodology for building the flow-through electrochemical filter.

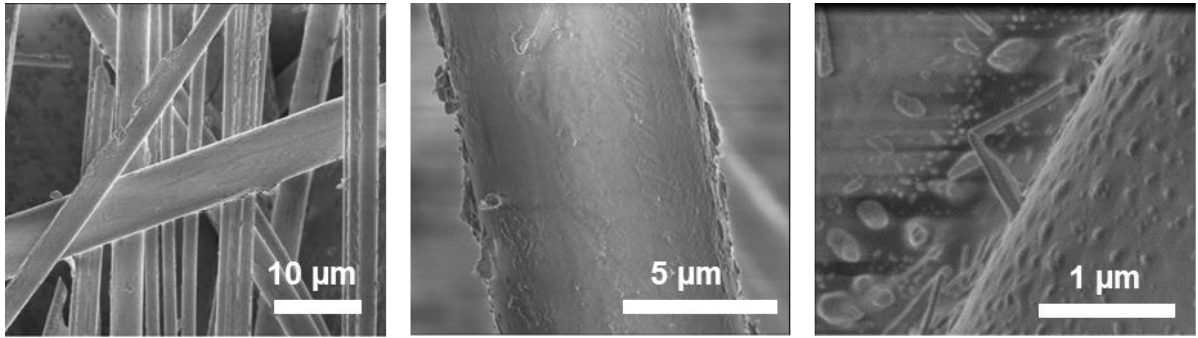

**Figure S2.** Scanning electron microscopy (SEM) images of uncoated mineral wool.

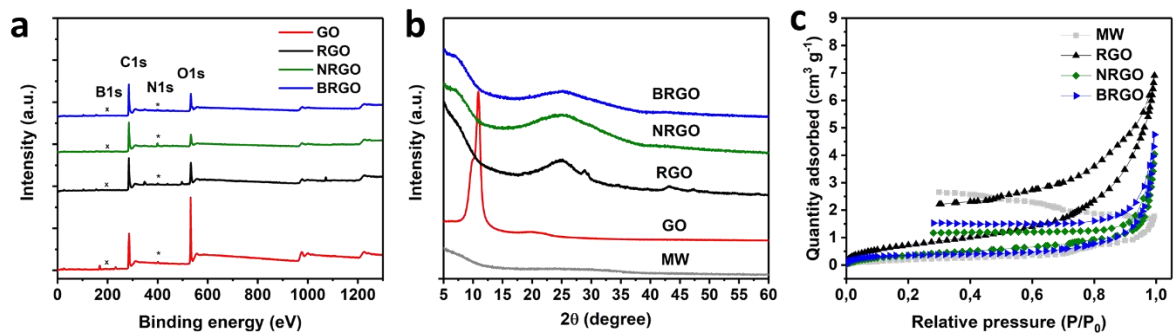

**Figure S3.** Characterization of the initial graphene oxide (GO), mineral wool and graphene-based sponges: **a)** X-ray photoelectron spectroscopy (XPS) spectra, **b)** X-ray powder diffraction (XRD) analyses and **c)** $N_2$ adsorption-desorption isotherms.

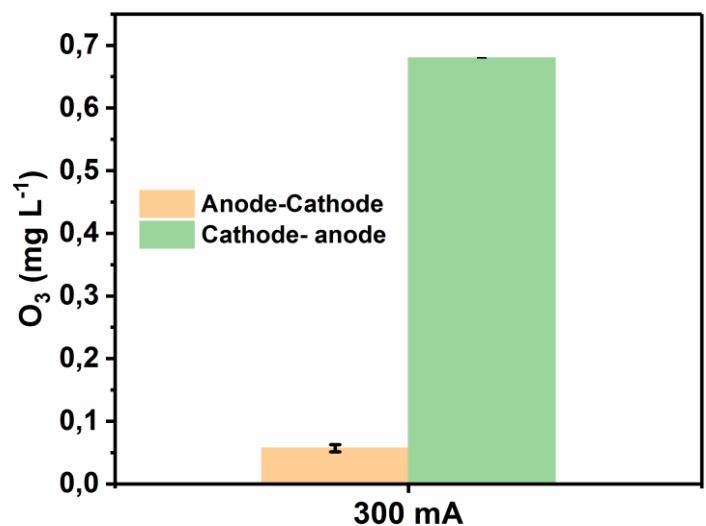

**Figure S4**. $O_3$ generation at anode-cathode and cathode-anode at 300 mA in 100 mM $KH_2PO_4/K_2HPO_4$ buffer (pH 7.2, 11 mS cm$^{-1}$). The values presented are mean values of the experiments conducted in triplicate, with their standard deviations.

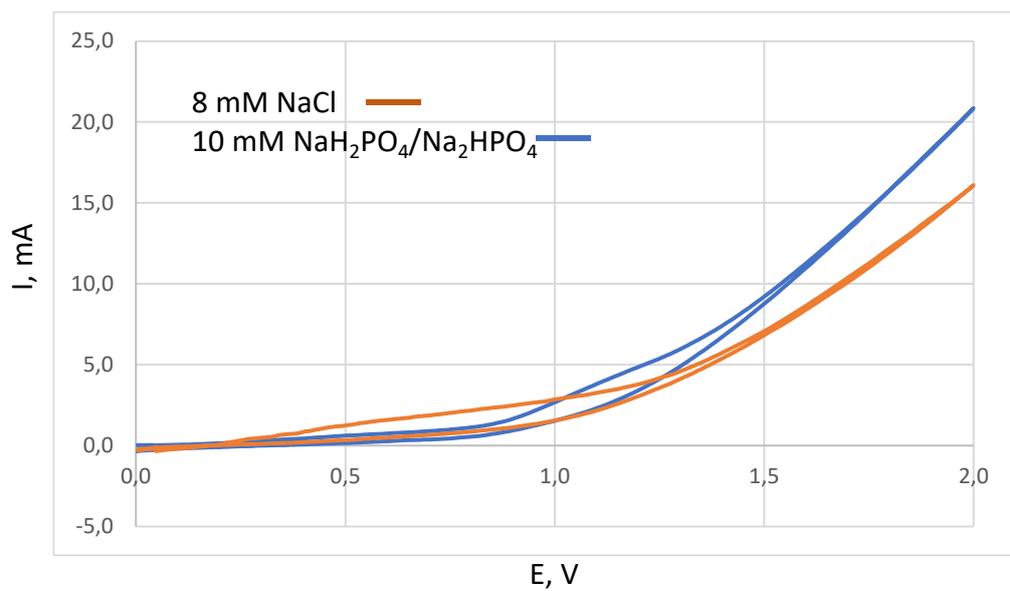

**Figure S5**. Cyclic voltammetry of the BRGO anode recorded in 10 mM $Na_2HPO_4/NaH_2PO_4$ buffer and in 8 mM NaCl solution, both at pH 7, 1.2 mS cm$^{-1}$, and using 10 mV s$^{-1}$ scan rate.

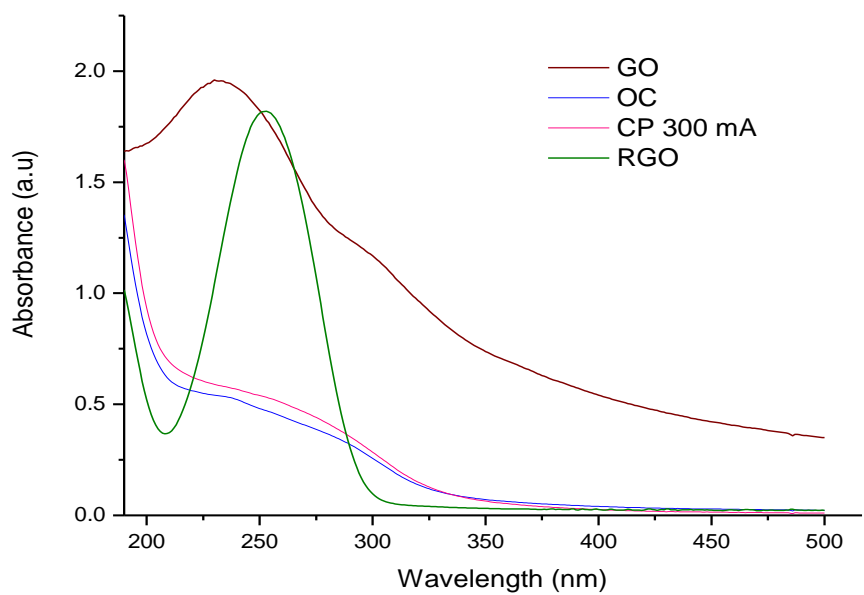

**Figure S6** Absorption spectra of: GO solution at 40 mg L$^{-1}$ (brown), RGO solution* at 40 mg L$^{-1}$ (green), and effluent samples taken after the open circuit (OC) (blue) and chronopotentiometric (CP) experiment at 300 mA (red). GO displays an absorption peak at 230 nm in accordance with the literature.

*To illustrate a shift in absorbance of GO to the higher wavelengths, RGO solution was prepared using milder reduction procedure to facilitate the re-suspension of RGO and absorbance measurement. Given that this reduction was incomplete, the absorbance peak was located at 254 nm, instead of 268-270 nm reported in literature [1,2].

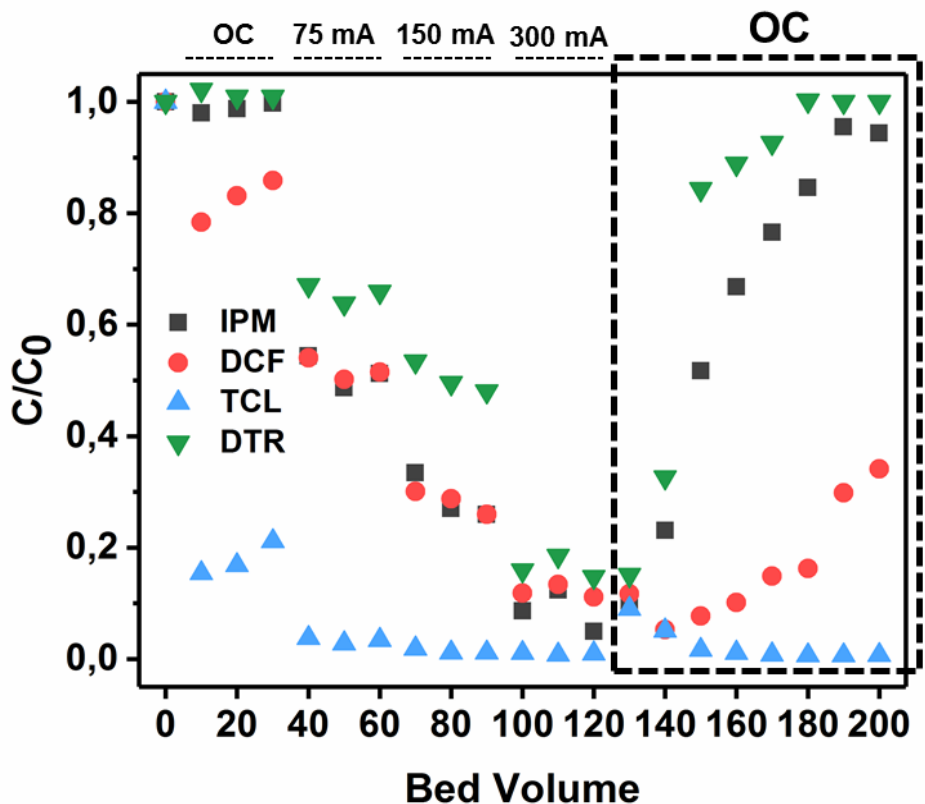

**Figure S7.** Effluent concentrations normalized to the feed concentrations (i.e., $C/C_0$) of target contaminants in the open circuit (OC) and during electrochemical treatment in anode-cathode configuration (i.e., BRGO-NRGO) at varying anodic current densities. After switching off the current the concentrations of contaminants increased but never above the feed concentration, thus further supporting their electrochemical degradation. In some cases (e.g., DCF) the permeate concentrations in the last OC experiment were lower compared with the first OC, which can be explained by the prolonged electrosorption of this compound due to the capacitance of graphene. IPM-iopromide, DCF-diclofenac, TCL-triclosan, DTR-diatrizoate. The values presented are mean values, whereas the obtained standard deviations were all lower than 0.05 and are not presented for the sake of clarity of the figure.

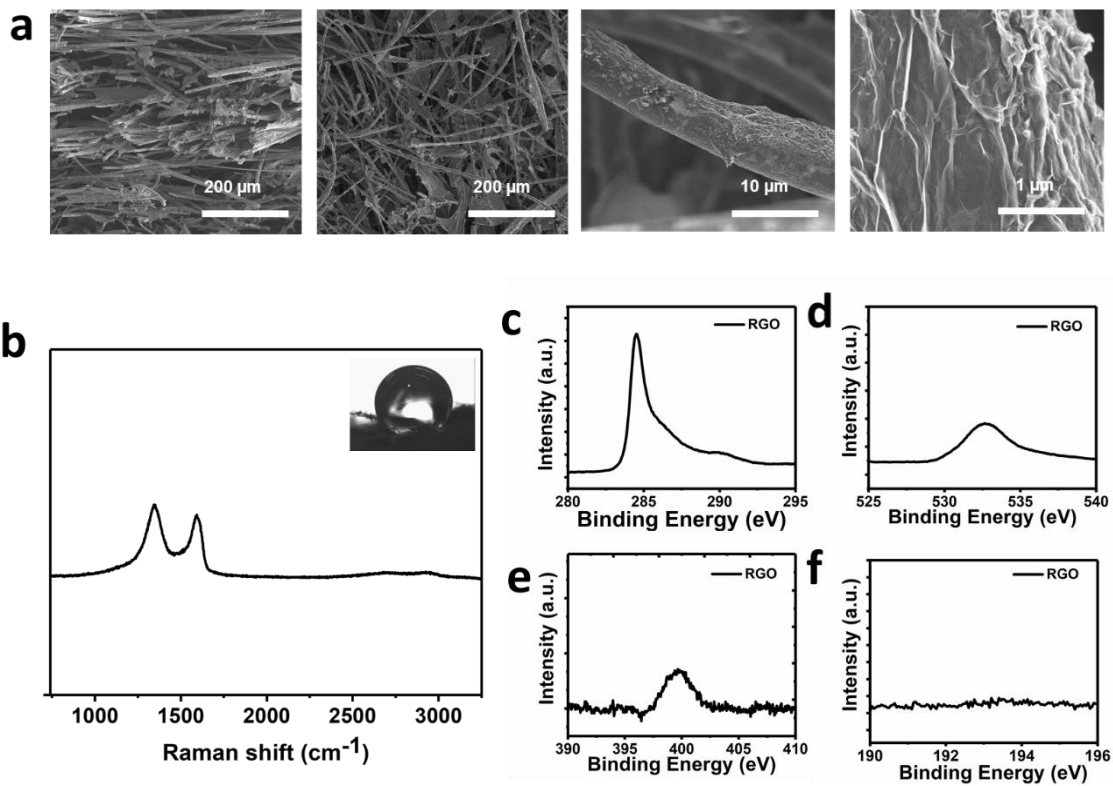

**Figure S8.** Optical characterization of RGO sponge: **a)** SEM images, **b)** Raman Spectroscopy profiles (inset: contact angle droplet over RGO), **c)** C1s, **d)** O1s, **e)** N1s and **f)** B1s XPS spectra.

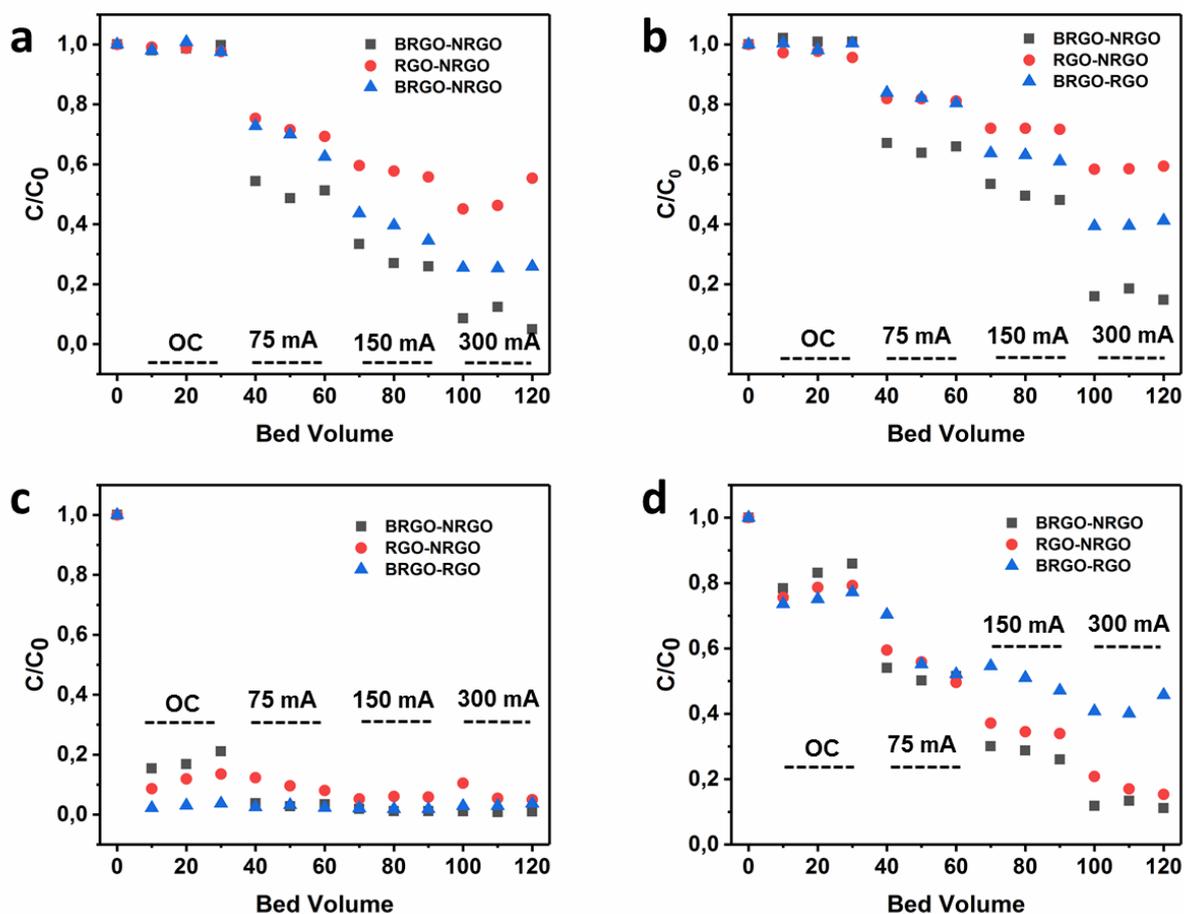

**Figure S9.** Electrochemical removal of persistent organic contaminants using graphene-sponge electrodes in anode-cathode flow direction in the open circuit (OC) and at different anodic currents and constant flow rate of 5 mL min$^{-1}$, using BRGO (A) – NRGO (C), RGO (A) – NRGO (C) and BRGO (A) – RGO (C) configurations; effluent concentrations (C) of target contaminants are normalized to their feed concentrations ($C_0$): **a)** IPM, **b)** DTR, **c)** TCL and **d)** DCF. The values presented are mean values, whereas the obtained standard deviations were all lower than 0.05.

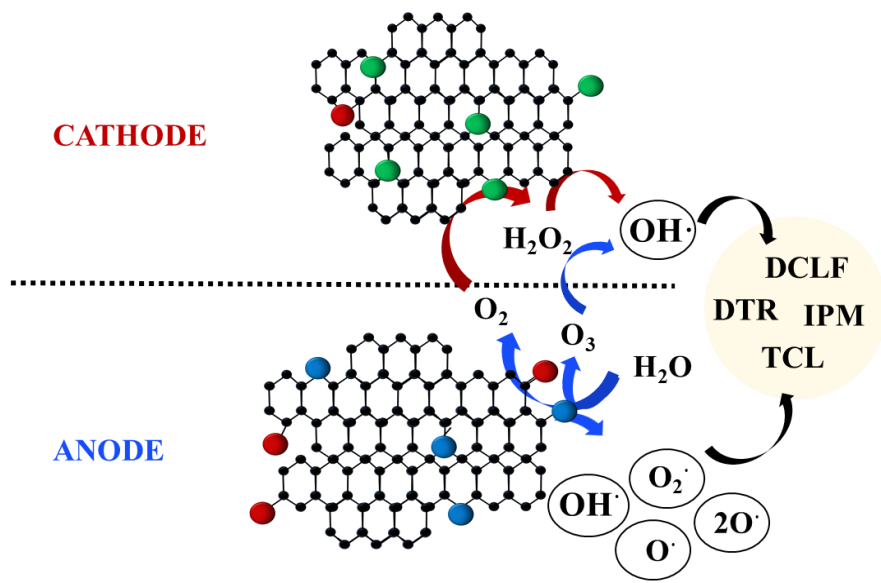

**Figure S10.** Schematic representation of the proposed degradation mechanism.

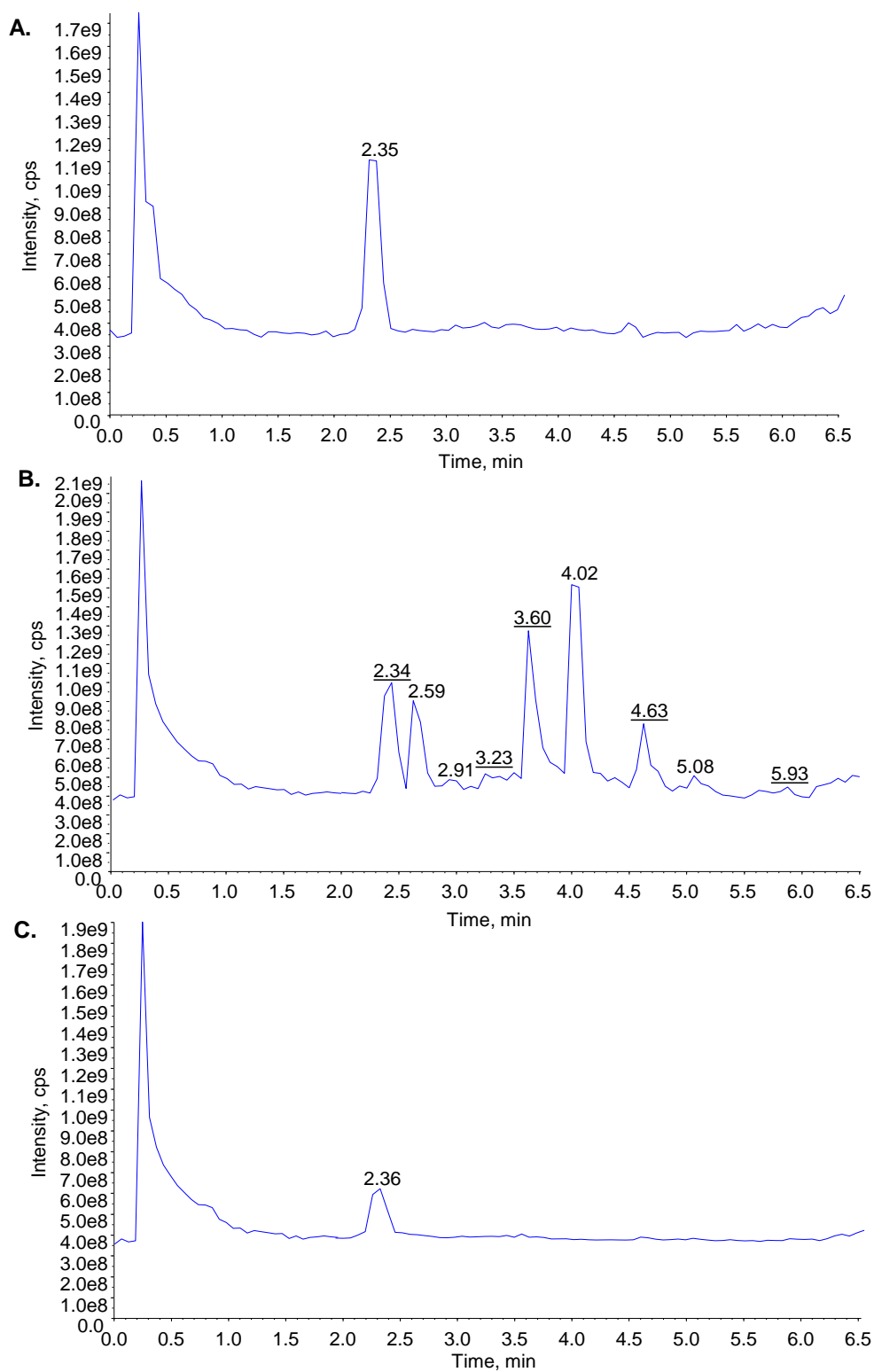

**Figure S11.** Total ion chromatograms (TICs) recorded in full scan mode for: **a)** sample at time, t=0, of the concentrated (25 µM) IPM solution, **b)** sample taken after the anode, at 30 bed volumes; the underlined retention times are those where molecular ions of TPs could be isolated from the full-scan, and **c)** treated sample (i.e., sample taken after the cathode) at 30 bed volumes.

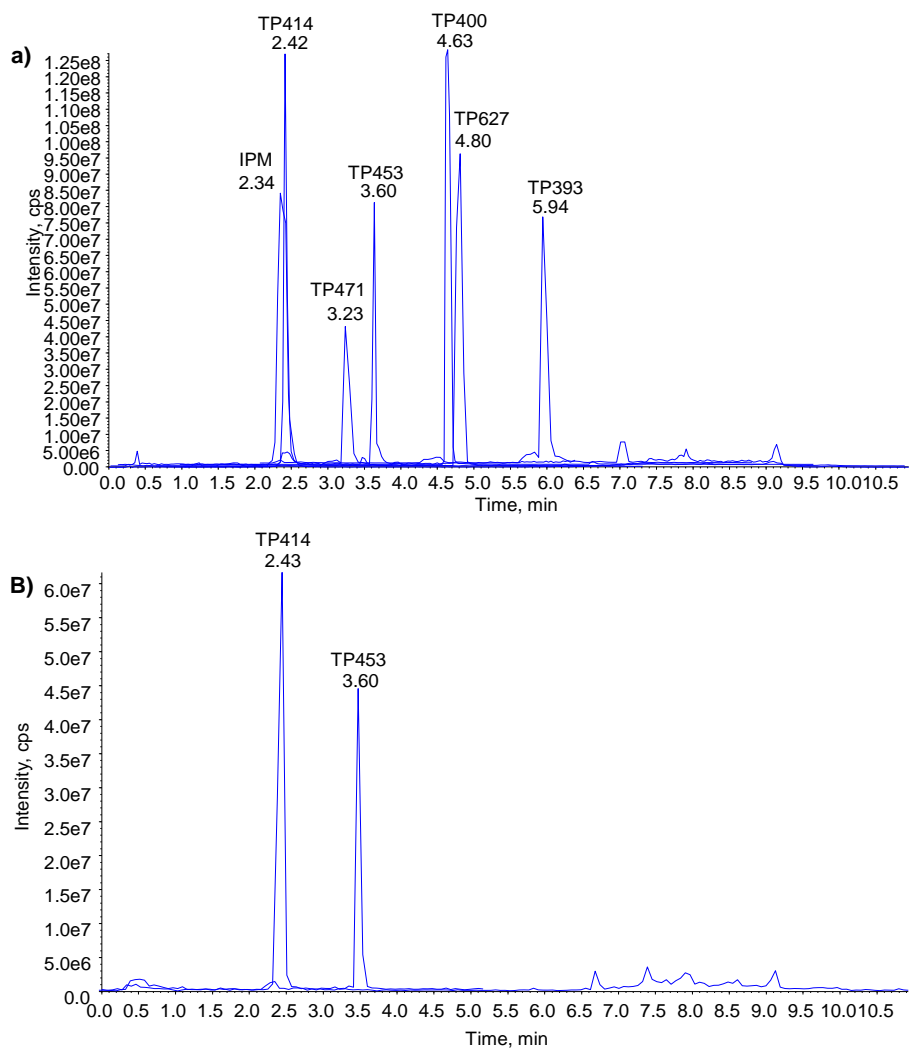

**Figure S12.** Extracted ion chromatograms (XICs) for IPM and its TPs with the identified molecular ions: **a)** sample taken after the anode, at 30 bed volumes, and **b)** treated sample (i.e., sample taken after the cathode) at 30 bed volumes.

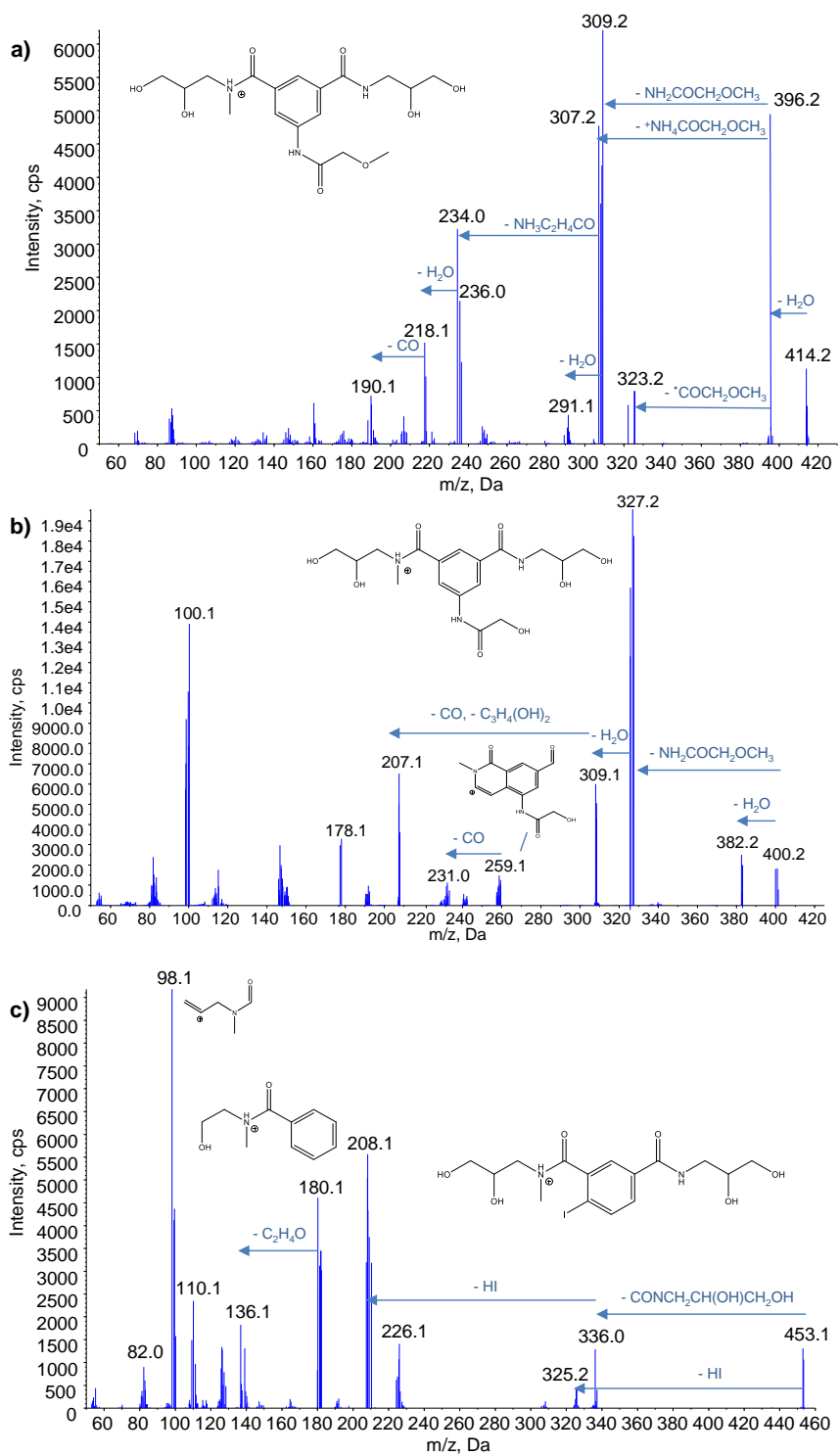

**Figure S13.** Collision induced dissociation (CID) MS$^2$ mass spectra and proposed fragmentation of the molecular ions for the identified TPs of IPM: **a)** TP414, [M+H]$^+$ 414.2, **b)** TP400, molecular ion [M+H]$^+$ 400.2, and **c)** TP453, [M+H]$^+$ 453.1.

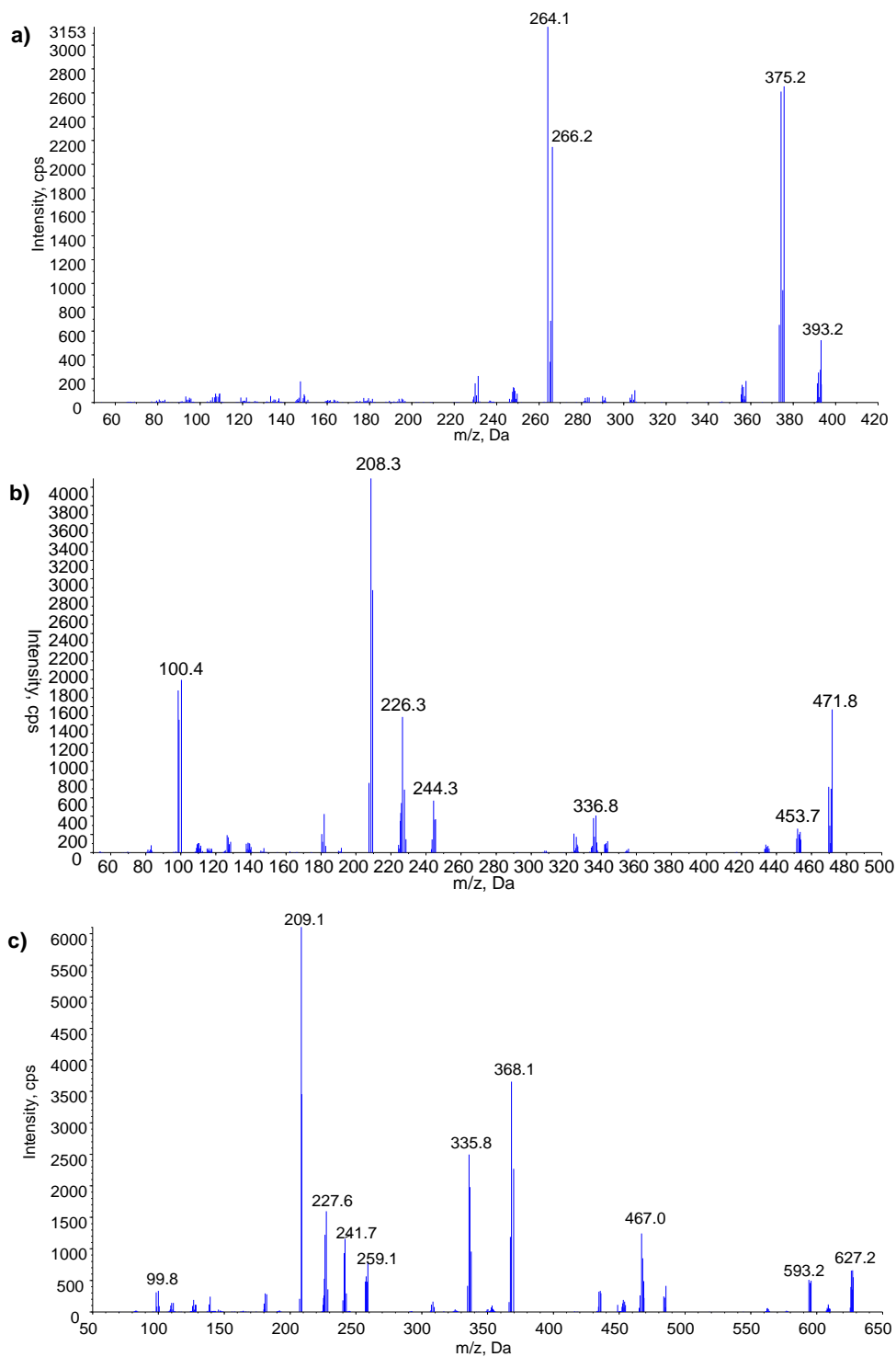

**Figure S14.** CID MS$^2$ mass spectra of the unidentified TPs of IPM: **a)** TP393, molecular ion [M+H]$^+$ 393.2, **b)** TP471, [M+H]$^+$ 471.8, and **c)** TP627, [M+H]$^+$ 627.2.

**Table S1.** The optimized compound-dependent MS parameters: declustering potential (DP), collision energy (CE) and cell exit potential (CXP) for each compound and each transition of the negative and positive mode.

| Organic compound | Q1 Mass (Da) | Q3 Mass (Da) | DP | CE | CXP |
|---|---|---|---|---|---|
| **Iopromide** | 791.72 | 572.9 | 156 | 35 | 20 |
| | 791.72 | 300.0 | 156 | 83 | 10 |
| **Diatrizoate** | 614.9 | 361 | 80 | 30 | 10 |
| | 614.9 | 233.1 | 85 | 33 | 10 |
| **Triclosan** | 287.0 | 35.4 | -40 | -35 | -7 |
| | 289.0 | 37.4 | -40 | -35 | -7 |
| **Diclofenac** | 293.92 | 250.0 | -65 | -16 | -11 |
| | 293.92 | 214.1 | -65 | -28 | -11 |

**Table S2.** XPS atomic content of GO precursor solution and synthesized graphene-based sponges.

|       | GO   | RGO  | NRGO | BRGO |
|-------|------|------|------|------|
| **C (%)** | 62.6 | 70.3 | 77.3 | 75.6 |
| **O (%)** | 36.9 | 28   | 16.1 | 21.8 |
| **N (%)** | 0.9  | 1.7  | 6.4  | 1.2  |
| **B (%)** | 0    | 0    | 0    | 1.3  |

**Table S3.** XPS Atomic content of the GO precursor solution and the synthesized graphene-based sponges. Percentage of functional groups of C1s, O1s and N1s XPS spectra of the synthetized graphene-based materials.

|  | C1s (%) | | | | O1s (%) | | | N1s (%) | | | |
| --- | --- | --- | --- | --- | --- | --- | --- | --- | --- | --- | --- |
|  | *C-C 284.5eV* | *C-N 285.6eV* | *C-O 286.9eV* | *COOH 288.5eV* | *C-O 531eV* | *C=O/ C-N-O 533eV* | *536.3eV* | *Pyridinic N 398.4eV* | *Pyrrolic N 399.8eV* | *Graphitic-N 401.6eV* | *Azide 402.7eV* |
| **GO** | 26.3 | 0 | 31.9 | 4.4 | 35.4 | 23.2 | 0 | 0 | 0 | 0.1 | 0 |
| **RGO** | 30.3 | 30.9 | 9.8 | 0 | 7.,8 | 16.5 | 3.6 | 0.4 | 0.7 | 0.,6 | 0 |
| **NRGO** | 24.7 | 37.1 | 10.8 | 4.6 | 9.8 | 6.1 | 0 | 2.5 | 3.0 | 0.4 | 0.4 |
| **BRGO** | 38.6 | 31.7 | 6.0 | 0 | 8.1 | 8.3 | 0 | 0.2 | 0.5 | 0.3 | 0.2 |

**Table S4.** Recorded anode ($E_{AN}$, V vs Standard Hydrogen Electrode, /SHE) and cathode potentials ($E_{CAT}$, V/SHE), and total cell potentials ($E_{TOT}$, V) at different currents for the four electrochemical systems.

|  | 75mA | | | 150mA | | | 300mA | | |
|---|---|---|---|---|---|---|---|---|---|
|  | $E_{AN}$, V/SHE | $E_{CAT}$, V/SHE | $E_{TOT}$, V | $E_{AN}$, V/SHE | $E_{CAT}$, V/SHE | $E_{TOT}$, V | $E_{AN}$, V/SHE | $E_{CAT}$, V/SHE | $E_{TOT}$, V |
| **RGO(A)-RGO(C)** | 2.6±0.4 | -4.5±0.3 | **7.1±0.1** | 3.4±0.1 | -5.9±0.5 | **9.5±0.7** | 5.2±1.2 | -16.9±0.9 | **22.2±0.3** |
| **BRGO(A)-RGO(C)** | 4.5±1.3 | -2.1±0.0 | **6.5±1.3** | 7.4±0.3 | -6.2±3.0 | **13.6±3.3** | 10.8±0.5 | -12.6±4.1 | **23.4±3.7** |
| **RGO(A)-NRGO(C)** | 3.2±0.5 | -2.3±0.3 | **5.5±0.2** | 5.1±0.1 | -3.3±0.4 | **8.4±0.4** | 7.6±1.4 | -4.30±1.3 | **11.9±0.1** |
| **BRGO(A)-NRGO(C)** | 2.1±0.1 | -3.0±0.3 | **5.2±0.2** | 3.8±0.1 | -4.2±0.1 | **8.03±0.2** | 5.7±0.0 | -6.1±0.3 | **11.9±0.3** |
| **NRGO(C)-BRGO(A)** | 2.1±0.1 | -2.6±0.1 | **4.7±0.0** | 3.1±0.3 | -2.9±0.9 | **6.1±0.2** | 4.2±0.6 | -3.5±0.9 | **7.7±0.2** |

**Table S5.** Chemical structures and physico-chemical properties of the target contaminants; molecular weight (MW), pKa, octanol-water distribution coefficient calculated based on chemical structure at pH 7.4 (ACD/logD), and polar surface area. Calculated ACD/logD values and polar surface areas were collected from Chemspider.com database.

| Organic compound (MW, g mol$^{-1}$) | Chemical structure | pKa | ACD/logD | Polar Surface area (Å$^2$)[3] |
|---|---|---|---|---|
| Iopromide (791.11) | | 11.1[4] | -2.12 | 169 |
| Diatrizoate (613.91) | | 2.2[4] | -2.12 | 96 |
| Triclosan (289.54) | | 7.9[5] | 5.13 | 29 |
| Diclofenac (318.13) | | 4.1[6] | 1.37 | 49 |

[3] Chemspider.com database (www.chemspider.com).
[4] https://www.drugbank.ca/
[5] Yoon, Y.; Westerhoff, P.; Snyder, S. A.; Wert, E. C.; Yoon, J. Removal of endocrine disrupting compounds and pharmaceuticals by nanofiltration and ultrafiltration membranes. Desalination 2007, 202 (1-3), 16-23. DOI: 10.1016/j.desal.2005.12.033.
[6] Verliefde, A. R. D.; Heijman, S. G. J.; Cornelissen, E. R.; Amy, G.; Van der Bruggen, B.; van Dijk, J. C. Influence of electrostatic interactions on the rejection with NF and assessment of the removal efficiency during NF/GAC treatment of pharmaceutically active compounds in surface water. Water Res.2007, 41 (15), 3227-3240. DOI: 10.1016/j.watres.2007.05.022.

**Table S6**. Reported bimolecular rate constants of target contaminants.

| Compound | k$_{O3}$ (M$^{-1}$ s$^{-1}$) | k$_{OH}$ (M$^{-1}$ s$^{-1}$) |
|---|---|---|
| Iopromide (IPM) | 93.1 [7] | 3.3±0.1×10$^9$ [8] |
| Diatrizoate (DTR) | 48.6 [7] | 9.6±0.2×10$^8$ [8] |
| Triclosan (TCL) | ᵃ3.8±0.1×10⁷ [9] | 5.4±0.1×10⁹ [10] |
| Diclofenac (DCF) | 6.8×10⁵ [11] | 9.3±0.1×10⁹ [12] |

[a] value given for anionic form of TCL present at pH 7.

**Text S1: Characterization of graphene-based sponge electrodes**

SEM was performed using a FEI Quanta FEG (pressure: 70Pa; HV: 20kV; and spot: four). Dispersive spectrometer Jobin-Yvon LabRam HR 800, coupled to an optical microscope Olympus BXFM was used for Raman characterization. The CCD detector was cooled at -70°C and a 532 nm laser line was used with a dispersive grating of 600 lines/mm and a laser power at sample of 0.5 mW. The $I_d/I_g$ ratios were calculated analyzing three different spots/areas per sample. XRD data was acquired with an X'pert multipurpose diffractometer at room temperature using a Cu Kα radiation (l = 1.540 Å). This has a vertical θ–θ goniometer (240 mm radius), where the sample stages are fixed and do not rotate around the Ω axis as in Ω–2θ diffractometers. The detector used is an X'Celerator that is an ultrafast X-ray detector based on a real time multiple strip technology. The diffraction pattern was recorded between 4° and 30° using a step size of 0.03° and a time per step of 1,000 s. The XPS measurements were done with a Phoibos 150 analyzer (SPECS GmbH, Berlin, Germany) in ultra-high vacuum conditions (base pressure $1^{-10}$ mbar) with a monochromatic aluminium Kalpha x-ray source (1486.74 eV). The energy resolution as measured by the FWHM of the Ag 3d5/2 peak for a sputtered silver foil was 0.58 eV. The Brunauer–Emmett–Teller (BET) specific surface area and pore size distribution were determined by $N_2$ adsorption-desorption at 77 K in (ASAP 2420, Micromeritics). The measurements of the contact angle of the synthesized graphene-based sponges were performed using EasyDrop Contact Angle Measuring Instrument by KRUSS GmbH in three different layers (one outer and two inner surfaces). The cross-plane electrical resistances of the graphene-based sponges were determined in a 0,5x0,5x0,5cm cube (x,y,z directions) between the two geometrically opposite faces.

**Text S2: Calculation of Ohmic drop**

The ohmic drop was calculated from the Ohmic internal resistance obtained in the electrochemical impedance spectroscopy (EIS) experiments of the BRGO anode:

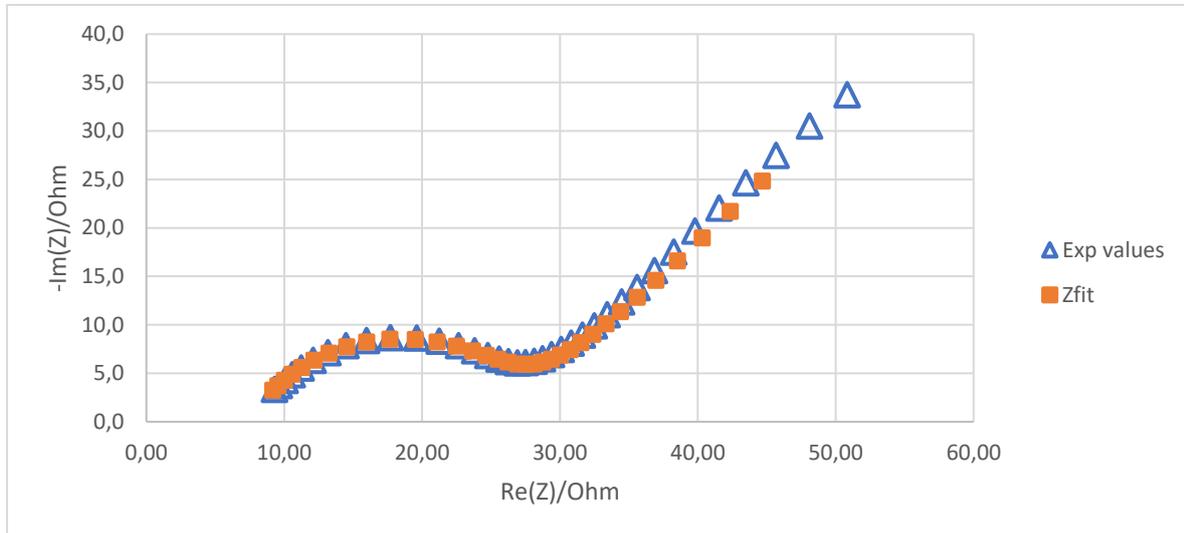

The EIS experimental data was fitted using the BioLogic EC-lab software using the equivalent circuit illustrated below:

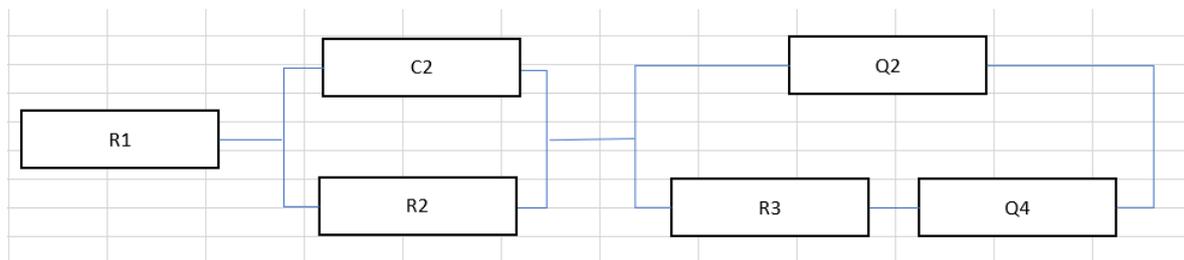

In the circuit, the Ohmic internal resistance (R1) is connected in series with the capacitance C2 and the charge transfer resistance R2 of the graphene sponge that are connected in parallel to each other. A constant phase element (Q2), representing the double layer capacitance, which occurs at the interface between the material and the electrolytes due to charge separation, is connected in parallel with a charge separation resistance R3, and Q4 that represents the ideal polarizable capacitance which would give rise to a straight line parallel to the imaginary axis. The obtained R1 was 6.425 Ω, thus resulting in the ohmic drop of 0.48 V, 0.96 V, and 1.9 V for 75 mA, 150 mA and 300 mA of the applied anodic current, respectively. The obtained $X^2$ was 25.12.

**Text S3: Analysis of target organic contaminants**

IPM and DTR were analyzed in electrospray (ESI) positive mode using an Acquity ultraperformance liquid chromatography (UPLC) HSS T3 column (2.1×50 mm, 1.8 μm, Waters) run at 30°C. The eluents employed were acetonitrile with 0.1% formic acid (eluent A), and milli-Q (LC-MS grade) water with 0.1% formic acid (eluent B) at a flow rate of 0.5 mL $min^{-1}$. The gradient was started at 2% of eluent A that was increased to 20% A by 3 min, further increased to 50% A by 6 min and further increased to 95% A by 7 min. It was kept constant for 2.5 min, before returning to the initial condition of 2% A by 9.5 min. The total run time was 11 min. TCL and DCF were analyzed in the ESI negative mode using an Acquity UPLC© BEH C18 column (2.1×50mm, 1.7μm) from Waters run at 30°C. The eluents for ESI negative mode were milli-Q (LC-MS grade) water containing a mixture of acetonitrile and methanol (1:1, v/v) (eluent A) and 1 mM ammonium acetate (eluent B) at a flow rate of 0.6 mL $min^{-1}$. The gradient was started at 5% A, further increased to 100% A by 7 min and then kept constant for 2 min, before returning to the initial conditions of 5 % A by 9 min. The total run time in the ESI negative mode was 11 min.

The target organic contaminants were analyzed in a multiple reaction monitoring (MRM). The settings for the compound-dependent parameters of each transition are summarized in **Table S1**. The source-dependent parameters were as follows: for the positive mode; curtain gas (CUR), 30 V; nitrogen collision gas (CAD), medium; source temperature (TEM), 650°C; ion source gases GS1, 60 V and GS2, 50 V; ion spray voltage, 5500V, and entrance potential (EP), 10V. For the negative mode; curtain gas (CUR), 30V; nitrogen collision gas (CAD), medium;

source temperature (TEM), 650°C; ion source gases GS1, 60 V and GS2, 70 V; ion spray voltage, -3500 V, and entrance potential (EP), -10V.

Tentative identification of TPs of IPM was performed by full-scan mode of analysis, isolation of the protonated molecular ions, collision induced dissociation (CID) $MS^2$ experiments in (+)ESI mode and mass spectral comparison with the parent compound, as well as with the literature data.

**Text S4: Current efficiency**

Current efficiencies (CE. %) for anodic reactions were calculated from the formula (12.13):

$$CE = n * F * q * \frac{(C_t - C_o)}{I} * 100, \ \% \quad \text{(eq. S1)}$$

where $n$ is the number of electrons involved in the reaction ($F$ is Faraday constant (96.485 C mol$^{-1}$). $q$ is applied flowrate of supporting electrolyte (L s$^{-1}$). $I$ is applied current (A). and $C_0$ and $C_t$ (mol L$^{-1}$) are initial, and concentration of a given species measured at time $t$.

In the case of anodic formation of chlorine. CE was calculated using $n=2$. based on the equation:

$$2Cl^- \rightarrow Cl_2 + 2e^-. \ E°=1.36 \ \text{V/SHE}. \quad \text{(eq. S2)}$$

In the case of anodic formation of ozone. CE was calculated using $n=8$. based on the equations:

$$3H_2O \rightarrow O_3 + 6H^+ + 6e^-. \ E°=1.51 \ \text{V/SHE}. \quad \text{(eq. S3)}$$

$$O_2 + H_2O \rightarrow O_3 + 2H^+ + 2e^-. \ E°=2.07 \ \text{V/SHE}. \quad \text{(eq. S4)}$$

Ozone gas evolves by a six-electron reaction in Eq. S2 at a voltage higher than 1.51 V/SHE. accompanied by oxygen evolution. By increasing voltage to above 2.07 V/SHE. the oxidation of O$_2$ gas to form O$_3$ is also expected as shown in Eq. S3.

The removal of persistent organic contaminants was expressed using the electrooxidation flux ($J_{ELOX}$, mol m$^{-2}$ h$^{-1}$), calculated according to the following:

$$J_{ELOX} = (C_0 - C) \times J \quad \text{(eq. S5)}$$

where $C_0$ and $C$ are feed and effluent concentrations of a given contaminant (µM), respectively, and $J$ is the effluent flux (L m$^{-2}$ h$^{-1}$, LMH) calculated by dividing the volumetric flow rate over the projected surface area of the electrodes (17.34 cm$^2$). The electric energy consumption (Ec, kWh m$^{-3}$) was calculated according to the following equation:

$$E_c = \frac{U \cdot I}{q \cdot \log\frac{C_0}{C}} \qquad (eq.\ S6)$$

where $q$ is the flow rate (L h$^{-1}$), $U$ the average cell voltage (V) and $I$ the applied current (A). Note that $\log(C_0/C)=1$ for one order of magnitude removal, and $\log(C_0/C)<1$ for removals lower than 90%. All experiments and measurements were performed in triplicate and the results were expressed as mean values with their standard deviations.

**Text S5. Characterization of the undoped RGO sponge.**

The SEM analysis of the graphene sponge demonstrated the coating of the mineral wool template with graphene (**Figure S7a**). Based on the XPS analysis, the C/O atomic ratio of the RGO sponge was determined to be 2.51, thus indicating a lower reduction degree compared with B-GS (C/O ratio of 3.46) and N-GS (C/O ratio of 4.80) (**Figure S7c-f**). The undoped GS also had ~1% nitrogen atom content originating from the commercial GO solution employed and was not a result of the reduction methodology (**Figure S3a and S7c-f**). The XRD analysis showed a d-spacing of ≈ 3.51 Å (**Figure S3b**). The Raman Spectroscopy determined an $I_d/I_g$ ratio of 1.15 (**Figure S7b**). The BET specific surface area of GS was 1.39 m$^2$ g$^{-1}$ (**Figure S3c**), and its contact angle was 137.3°±3.4° (inset in **Figure S7b**). The undoped graphene-based sponge electrode had a through-plane resistance of 32.6±18.6 kΩ and in-plane resistance of 5.2±3.8 kΩ.